\documentclass[twocolumn,nofootinbib,superscriptaddress,preprintnumbers,floatfix]{revtex4}

\usepackage[bookmarks=false,hyperfootnotes=false]{hyperref}
\usepackage{graphicx}
\usepackage{amsmath}

\newcommand{\eq}[1]{Eq.~\eqref{eq:#1}}
\newcommand{\eqs}[2]{Eqs.~\eqref{eq:#1} and \eqref{eq:#2}}
\renewcommand{\sec}[1]{Sec.~\ref{sec:#1}}
\newcommand{\subsec}[1]{Sec.~\ref{subsec:#1}}
\newcommand{\fig}[1]{Fig.~\ref{fig:#1}}

\newcommand{\abs}[1]{\lvert#1\rvert}
\newcommand{\ord}[1]{\mathcal{O}(#1)}

\newcommand{\df}{\mathrm{d}}

\newcommand{\nb}{\,\mathrm{nb}}
\newcommand{\pb}{\,\mathrm{pb}}
\newcommand{\GeV}{\,\mathrm{GeV}}
\newcommand{\TeV}{\,\mathrm{TeV}}

\newcommand{\nn}{\nonumber}

\newcommand{\tot}{\mathrm{total}}
\newcommand{\cut}{\mathrm{cut}}
\newcommand{\jet}{\mathrm{jet}}

\newcommand{\Ecm}{{E_\mathrm{cm}}}
\newcommand{\Tcm}{{\mathcal{T}_\mathrm{cm}}}
\newcommand{\Tcmc}{\mathcal{T}_\mathrm{cm}^\mathrm{cut}}

\allowdisplaybreaks[4]

\begin{document}

%%%%%%%%%%%%%%%%%%%%%%%%%%%%%%%%%%%%%%%%%%%%%%%%%%%%%%%%%%%%%%%%%%%%%%%%%%%%%%%%
% Title page
%%%%%%%%%%%%%%%%%%%%%%%%%%%%%%%%%%%%%%%%%%%%%%%%%%%%%%%%%%%%%%%%%%%%%%%%%%%%%%%%

\preprint{\vbox{\hbox{MIT--CTP 4281}}}

\title{Theory Uncertainties for Higgs and Other Searches Using Jet Bins}

\author{Iain W.~Stewart}
\affiliation{Center for Theoretical Physics, Massachusetts Institute of Technology,
Cambridge, Massachusetts 02139, USA\vspace{0.5ex}}
\affiliation{Center for the Fundamental Laws of Nature, Harvard University,
Cambridge, Massachusetts 02138, USA\vspace{0.5ex}}

\author{Frank J.~Tackmann}
\affiliation{Center for Theoretical Physics, Massachusetts Institute of Technology,
Cambridge, Massachusetts 02139, USA\vspace{0.5ex}}

\date{July 11, 2011}

%%%%%%%%%%%%%%%%%%%%%%%%%%%%%%%%%%%%%%%%%%%%%%%%%%%%%%%%%%%%%%%%%%%%%%%%%%%%%%%%
\begin{abstract}

  Bounds on the Higgs mass from the Tevatron and LHC are determined using
  exclusive jet bins to maximize sensitivity.  Scale variation in exclusive
  fixed-order predictions underestimates the perturbative uncertainty for these
  cross sections, due to cancellations between the perturbative corrections
  leading to large $K$ factors and those that induce logarithmic sensitivity to
  the jet-bin boundary.  To account for this, we propose that scale variation in
  the fixed-order calculations should be used to determine theory uncertainties
  for inclusive jet cross sections, whose differences yield exclusive jet cross
  sections.  This yields a theory correlation matrix for the jet bins such that
  the additional uncertainty from large logarithms due to the jet boundary
  cancels when neighboring bins are added.  This procedure is tested for $H+0,1$
  jets, $WW+ 0$ jets, and $W+0,1,2$ jets, and found to be generally applicable.
  For a case where the higher-order resummation of the jet boundary corrections
  is known, we show that this procedure yields fixed-order uncertainties which
  are theoretically consistent with those obtained in the resummed calculation.

\end{abstract}
%%%%%%%%%%%%%%%%%%%%%%%%%%%%%%%%%%%%%%%%%%%%%%%%%%%%%%%%%%%%%%%%%%%%%%%%%%%%%%%%

\maketitle

%%%%%%%%%%%%%%%%%%%%%%%%%%%%%%%%%%%%%%%%%%%%%%%%%%%%%%%%%%%%%%%%%%%%%%%%%%%%%%%%
\section{Introduction}
\label{sec:intro}
%%%%%%%%%%%%%%%%%%%%%%%%%%%%%%%%%%%%%%%%%%%%%%%%%%%%%%%%%%%%%%%%%%%%%%%%%%%%%%%%

In the search for the Higgs boson at the Tevatron and the Large Hadron Collider
(LHC), the data are divided into exclusive jet bins. This is done because the
background composition depends on the number of jets in the final state, and
the overall sensitivity can be increased significantly by optimizing the
analysis for Higgs + $0$, $1$, and $2$ jet signals.
The primary example is the $H\to WW^*$ decay channel, which dominates the current
Tevatron exclusion limits around $m_H \simeq 2 m_W$~\cite{Aaltonen:2010yv, Aaltonen:2011gs}, and
is one of the important channels for $m_H \gtrsim 130\GeV$ being pursued at the
LHC~\cite{Chatrchyan:2011tz, Aad:2011kk}. The importance of the
Higgs + $1$ jet channel in $H\to \tau\tau$ and $H\to WW^*$ was demonstrated explicitly
in Refs.~\cite{Mellado:2004tj, Mellado:2007fb}. Similarly, for
$H\to\gamma\gamma$, which plays an important role for $m_H \lesssim 130\GeV$,
the search sensitivity can be improved by optimizing the analysis for different
jet bins~\cite{Aad:2009wy}.

Since the measurements are performed in each jet bin, the perturbative
uncertainties in the theoretical predictions must also be evaluated separately
for each jet multiplicity~\cite{Anastasiou:2009bt}. Furthermore, to combine the
results in the end, the correlations between the theoretical uncertainties in
the different jet bins as well as in the total cross section have to be
understood and taken into account.

In the winter 2011 Tevatron analyses of $gg\to H\to
WW^*$~\cite{Aaltonen:2011gs}, the perturbative uncertainties in the signal cross
section are evaluated using common scale variation for the exclusive jet bins,
which yields~\cite{Anastasiou:2009bt}
%%%
\begin{align} \label{eq:Tevunc}
\frac{\Delta\sigma_\tot}{\sigma_\tot} &=
66.5\% \times\! \Bigl({}^{+5\%}_{-9\%} \Bigr) + 28.6\% \times\! \Bigl({}^{+24\%}_{-22\%} \Bigr)
\nn\\ & \quad
+ 4.9\% \times\! \Bigl({}^{+78\%}_{-41\%}\Bigr) = \Bigl({}^{+14\%}_{-14\%}\Bigr)
\,.\end{align}
%%%
The three terms are the contributions from the $0$, $1$, and $(\geq2)$-jet
bins with their relative scale uncertainties in brackets. By using a common
scale variation the uncertainties are effectively 100\% correlated and are added
linearly, such that the $\pm 14\%$ scale uncertainty in the total
cross section is reproduced.

For the $0$-jet bin, which is the most sensitive search channel in $H\to WW^*$,
one applies a strong veto on additional jets.  It is often argued that with the
jet veto the perturbative uncertainties improve, yielding scale uncertainties
from fixed-order perturbation theory that are smaller than those in the total
cross section, as seen in \eq{Tevunc}. This apparent improvement arises from
cancellations between two sources, large corrections to the total cross section
(large $K$ factors) and the large corrections from logarithmic dependence on the
jet veto.  Since the improvement arises from a cancellation between two large
and predominantly independent perturbative series, it must be assessed
carefully.

We propose a simple procedure to estimate more realistic perturbative uncertainties
for exclusive jet bins from fixed-order perturbation theory. The method is
designed for processes with large $K$ factors or large perturbative corrections in
inclusive cross sections and takes into account the structure of the various perturbative
series. As we will see, it can also be applied in general.
The essential idea is to first independently determine the
uncertainties in the inclusive $N$-jet cross sections $\sigma_{\ge N}$, and then
use them to compute the uncertainty in the exclusive $N$-jet cross section
$\sigma_N$ from the difference
%%%
\begin{equation} \label{eq:sigN}
\sigma_N = \sigma_{\ge N} - \sigma_{\geq N+1}
\,.\end{equation}
%%%
To a first approximation the perturbative series for $\sigma_{\ge N}$ can be
considered unrelated for different $N$. For instance, their series start at
different orders in $\alpha_s$, and there is a priori no direct relation between
the modifications to the series caused by the jet cuts that define these two
inclusive samples.  Therefore, as explained in detail in \sec{pert}, we can work in the limit
where the fixed-order perturbative uncertainties in the $\sigma_{\ge N}$'s can be taken as
uncorrelated, leading to
%%%
\begin{equation} \label{eq:DelN}
\Delta_N^2 = \Delta_{\geq N}^2 + \Delta_{\geq N+1}^2
\,.\end{equation}
%%%
The uncertainty in the exclusive cross section is larger than that in the
corresponding inclusive one, which accounts for its more complicated
perturbative structure. Equation~\eqref{eq:sigN} also leads to an anticorrelation
between the cross sections in neighboring jet bins. When neighboring
bins are added the sensitivity to the boundary between them cancels and the uncertainty
reduces accordingly.

For example, for the $0$-jet bin in $H\to WW^*$ discussed above, we have
$\sigma_0=\sigma_{\tot}-\sigma_{\geq 1}$. Here, $\sigma_{\geq 1}$ contains double logarithms
of the jet $p_T$ cut, whereas $\sigma_\tot$ does not involve any jet definition, so
their perturbative series can be considered largely independent. Therefore, taking their
perturbative uncertainties $\Delta_\tot$ and $\Delta_{\geq 1}$ as uncorrelated, the
covariance matrix for $\{\sigma_0, \sigma_{\ge 1}\}$ is%
\footnote{Since these are theory
  uncertainties, there is no strict reason to combine them in a particular way.
  We add them in quadrature since this is the most convenient for discussing
  correlations and error propagation.}
%%%
\begin{equation} \label{eq:01matrix}
\begin{pmatrix}
   \Delta_{\tot}^2 +\Delta_{\ge 1}^2 &  - \Delta_{\ge 1}^2 \\
  - \Delta_{\ge 1}^2 & \Delta_{\ge 1}^2
\end{pmatrix}\,.
\end{equation}
%%%
Using this matrix to compute the uncertainty in $\sigma_0+\sigma_{\ge 1}$
reproduces $\Delta_{\tot}$ as it should.

We should mention that we are only discussing here the uncertainties due to
unknown higher-order perturbative corrections, which are commonly estimated
using scale variations. We do not discuss parametric uncertainties, such as
parton distribution function (PDF) and $\alpha_s$ uncertainties,
which have been extensively discussed, recently for
example in Refs.~\cite{Baglio:2010um, Alekhin:2010dd, Dittmaier:2011ti,
  Baglio:2011wn, Alekhin:2011ey, Watt:2011kp, Thorne:2011kq, Baglio:2011hc}.

In the next section we present the arguments leading to our proposal for
evaluating the perturbative uncertainties for exclusive jet bins, and discuss
the structure of the perturbative series. In \sec{examples}, we apply our method
to a variety of processes. We start in Secs.~\ref{subsec:H0} and \ref{subsec:H1}
with discussion and numerical results for $gg\to H+0$ jets and $gg\to H+1$ jets.
In \subsec{WW}, we consider $pp\to WW+0$ jets, which is an important background
for Higgs production. In Secs.~\ref{subsec:W0}, \ref{subsec:W1},
and~\ref{subsec:W2} we consider $W+0,1,2$ jets, which are important backgrounds
for missing-energy searches. In \sec{resum}, we consider again $gg\to H+0$ jets
and test our method for the fixed-order uncertainties against a case where the
resummation of the large logarithms induced by the jet binning is known to
next-to-next-to-leading logarithmic (NNLL) accuracy.  We conclude in
\sec{conclusions}. In the Appendix, we give expressions for the
uncertainties and correlations for the case where one considers $0$, $1$, and
$(\geq 2)$-jet bins as in \eq{Tevunc}.

%%%%%%%%%%%%%%%%%%%%%%%%%%%%%%%%%%%%%%%%%%%%%%%%%%%%%%%%%%%%%%%%%%%%%%%%%%%%%%%%
\section{Jet Bin Uncertainties}
\label{sec:pert}
%%%%%%%%%%%%%%%%%%%%%%%%%%%%%%%%%%%%%%%%%%%%%%%%%%%%%%%%%%%%%%%%%%%%%%%%%%%%%%%%

To examine in more detail the modification of the perturbative series that takes place for
exclusive jet bins, we will consider the example of the $0$-jet bin and
$(\ge 1)$-jet bin.  The total cross section, $\sigma_\tot$, is divided
into a $0$-jet exclusive cross section, $\sigma_0(p^\cut)$, and the $(\geq 1)$-jet
inclusive cross section, $\sigma_{\geq 1}(p^\cut)$,
%%%
\begin{align} \label{eq:pcut}
\sigma_\tot &= \int_0^{p^\cut}\!\df p\, \frac{\df\sigma}{\df p} + \int_{p^\cut}\!\df p\, \frac{\df\sigma}{\df p}
\nn\\*
&\equiv  \sigma_0(p^\cut) + \sigma_{\geq 1}(p^\cut)
\,.\end{align}
%%%
Here, $p$ denotes the kinematic variable which is used to divide the cross
section into jet bins.  For most of our analysis we take $p \equiv p_T^\jet$,
which for \eq{pcut} is the largest $p_T$ of any jet in the event. In this case,
$\sigma_0(p_T^\cut)$ only contains events with jets having $p_T \leq p_T^\cut$,
and $\sigma_{\geq 1}(p_T^\cut)$ contains events with at least one jet with $p_T
\geq p_T^\cut$.

In \eq{pcut} both $\sigma_0$ and $\sigma_{\geq 1}$ depend on the phase space
cut, $p^\cut$, and by construction this dependence cancels in their sum.  This
means that the additional perturbative uncertainty induced by this cut, call it
$\Delta_\cut$, must be 100\% anticorrelated between $\sigma_0(p^\cut)$ and
$\sigma_{\geq 1}(p^\cut)$. That is, the contribution of $\Delta_\cut$ to the
covariance matrix for $\{\sigma_0, \sigma_{\geq 1}\}$ must be of the form
%%%
\begin{equation} \label{eq:cutmatrix}
C_\cut = \begin{pmatrix}
   \Delta_\cut^2 &  - \Delta_\cut^2 \\
   -\Delta_\cut^2 & \Delta_\cut^2
\end{pmatrix}\,.
\end{equation}
%%%
The questions then are: (1) how can we estimate $\Delta_\cut$, and (2) how is
the overall perturbative uncertainty $\Delta_\tot$ of $\sigma_\tot$ related to
the uncertainty for $\sigma_0$ and $\sigma_{\geq 1}$.

To answer these questions, we discuss the perturbative structure of
the cross sections in more detail. By restricting the cross section
to the $0$-jet region, one restricts the collinear initial-state radiation
from the colliding hard partons as well as the overall soft radiation in the
event. This restriction on additional emissions leads to the appearance of
Sudakov double logarithms of the form $L^2 = \ln^2(p^\cut/Q)$ at each order in
a perturbative expansion in the strong coupling constant
$\alpha_s$, where $Q$ is the hard scale of the process. For Higgs production
from gluon fusion, $Q = m_H$,
and the leading double logarithms appearing at $\ord{\alpha_s}$ are
%%%
\begin{align} \label{eq:sig0dbleL}
\sigma_0(p_T^\cut) &= \sigma_B \Bigl(1 - \frac{3\alpha_s}{\pi}\, 2\ln^2 \frac{p_T^\cut}{m_H} + \dotsb \Bigr)
\,,\end{align}
%%%
where $\sigma_B$ is the Born (tree-level) cross section.

The total cross section just depends on the hard scale $Q$,
which means by choosing the scale $\mu \simeq Q$, the
fixed-order expansion does not contain large logarithms and has the
structure\footnote{These expressions for the perturbative series are schematic.
  They do not show the convolution with the parton distribution functions
  contained in $\sigma_B$, nor do they display $\mu$ dependent logarithms.
  In particular, the single logarithms related to the PDF evolution are not
  displayed, since they are not the logarithms we are most interested in discussing. }
%%%
\begin{equation} \label{eq:sigmatot}
\sigma_\tot \simeq \sigma_B\big[ 1 + \alpha_s + \alpha_s^2 + \ord{\alpha_s^3} \big]
\,.\end{equation}
%%%
The coefficients of this series can be large, corresponding to the well-known
large $K$ factors. For instance, the cross section for $gg\to H$ doubles from leading order
to next-to-leading order (NLO) even though $\alpha_s\sim 0.1$. As usual, varying the scale in $\alpha_s$
(and the PDFs) one obtains an estimate of the size of the missing higher-order
terms in this series, corresponding to $\Delta_\tot$.

The inclusive $1$-jet cross section has the perturbative
structure
%%%
\begin{align} \label{eq:sigma1}
\sigma_{\geq 1}(p^\cut)
&\simeq \sigma_B\big[ \alpha_s (L^2 + L + 1)
\\* \nn & \quad
+ \alpha_s^2 (L^4 + L^3 + L^2 + L + 1) + \ord{\alpha_s^3 L^6} \big]
\,,\end{align}
%%%
where the logarithms $L = \ln(p^\cut/Q)$ arise from cutting off the IR
divergences in the real emission diagrams. For $p^\cut \ll Q$ the logarithms can
get large enough to overcome the $\alpha_s$ suppression. In the limit $\alpha_s
L^2 \simeq 1$, the fixed-order perturbative expansion breaks down and the
logarithmic terms must be resummed to all orders in $\alpha_s$ to obtain a
meaningful result. For typical experimental values of $p^\cut$ fixed-order
perturbation theory can still be considered, but the logarithms cause large
corrections at each order and dominate the series. This means varying the scale
in $\alpha_s$ in \eq{sigma1} directly tracks the size of the large logarithms
and therefore allows one to get some estimate of the size of missing
higher-order terms caused by $p^\cut$, that correspond to $\Delta_\cut$.
Therefore, we can approximate $\Delta_\cut = \Delta_{\geq 1}$, where
$\Delta_{\geq 1}$ is obtained from the scale variation for $\sigma_{\geq 1}$.

The exclusive $0$-jet cross section is equal to the
difference between \eqs{sigmatot}{sigma1}, and so has the schematic structure
%%%
\begin{align} \label{eq:sigma0}
\sigma_0(p^\cut)
&\simeq \sigma_B \Big\{ \bigl[ 1 + \alpha_s + \alpha_s^2 + \ord{\alpha_s^3} \bigr]
\nn\\ & \quad
- \bigl[\alpha_s (L^2 \!+ L + 1) + \alpha_s^2 (L^4 \!+ L^3 \!+ L^2 \!+ L + 1)
\nn\\ & \qquad\quad
+ \ord{\alpha_s^3 L^6} \bigr] \Big\}
\,.\end{align}
%%%
In this difference, the large positive corrections in $\sigma_\tot$ partly
cancel against the large negative logarithmic corrections. For example, at
$\ord{\alpha_s}$ there is a value of $L$ for which the $\alpha_s$ terms in the
schematic \eq{sigma0} cancel exactly, indicating that at this $p^\cut$ the NLO
cross section has vanishing scale dependence and is equal to the LO cross
section, $\sigma_0(p^\cut)=\sigma_B$. We will see this effect explicitly in our
examples below, using the complete perturbative expressions.  We will find that
this occurs for values of $p^\cut$ in the experimentally relevant region. Because
of this cancellation, a standard use of scale variation in \eq{sigma0} does not
actually probe the size of the logarithms, and thus is not suitable to estimate
$\Delta_\cut$.

Since $\Delta_\cut$ and $\Delta_\tot$ are by definition uncorrelated, by
associating $\Delta_\cut = \Delta_{\geq 1}$ we are effectively treating
the perturbative series for $\sigma_\tot$ and $\sigma_{\geq 1}$ as independent
with separate (uncorrelated) perturbative uncertainties. That is, considering
$\{\sigma_\tot, \sigma_{\geq 1}\}$, the covariance matrix is diagonal,
%%%
\begin{equation} \label{eq:diagmatrix}
\begin{pmatrix}
  \Delta_\tot^2 & 0 \\ 0 & \Delta_{\geq1}^2
\end{pmatrix}
\,.\end{equation}
%%%
This is consistent, since for small $p^\cut$ the two series have very different
structures. In particular, there is no reason to believe that the same
cancellations in $\sigma_0$ will persist at every order in perturbation theory
at a given $p^\cut$.

From \eq{diagmatrix} it follows that the perturbative uncertainty in
$\sigma_0(p^\cut)$ is given by $\Delta_\tot^2 + \Delta_{\geq 1}^2$, i.e., by summing
the inclusive cross section uncertainties in quadrature. It also follows that the complete
covariance matrix for the three%
\footnote{The fact that only two of three are independent is reflected in the
  matrix, i.e. any $2\times2$ submatrix can be used to derive the full $3\times
  3$ matrix using the relation $\sigma_\tot = \sigma_0 + \sigma_{\geq 1}$.}
quantities $\{\sigma_\tot, \sigma_0, \sigma_{\geq 1}\}$ is
%%%
\begin{equation} \label{eq:fullmatrix}
C = \begin{pmatrix}
   \Delta_\tot^2 & \Delta_\tot^2 & 0 \\
   \Delta_\tot^2 & \Delta_{\geq 1}^2 + \Delta_\tot^2 &  - \Delta_{\geq 1}^2 \\
   0 & -\Delta_{\geq 1}^2 & \Delta_{\geq 1}^2
\end{pmatrix}\,,
\end{equation}
%%%
where $\Delta_\tot$ and $\Delta_{\geq 1}$ are considered uncorrelated and are
evaluated by separately varying the scales in the fixed-order predictions for
$\sigma_\tot$ and $\sigma_{\geq 1}(p^\cut)$, respectively. The $\Delta_{\geq 1}$
contributions in the lower right $2\times2$ matrix for $\sigma_0$ and
$\sigma_{\geq1}$ are equivalent to \eq{cutmatrix} with $\Delta_\cut =
\Delta_{\geq 1}$. Note that in this $2\times 2$ space all of $\Delta_\tot$
occurs in the uncertainty for $\sigma_0$.  This is reasonable from the point of
view that $\sigma_0$ starts at the same order in $\alpha_s$ as $\sigma_\tot$ and
contains the same leading virtual corrections.

The limit $\Delta_\cut = \Delta_{\geq 1}$ which \eq{fullmatrix} is based on is
of course not exact but an approximation. However, the preceding arguments show
that it is a more reasonable starting point than using a common scale variation
for the different jet bins. The latter usually results in the cross sections
being 100\% correlated, as in \eq{Tevunc}, and in particular does not account
for the additional $p^\cut$ induced uncertainties. In our numerical examples
below, we will see that our method produces more sensible uncertainty estimates
for fixed-order predictions. In \sec{resum} we will compare the estimates from
our method with those obtained by an explicit resummation in the jet-veto
variable. This provides further evidence that our method gives consistent
uncertainty estimates. Resummation provides a way for improving predictions for
the central value of the cross section, together with better estimates of
$\Delta_\cut$ and the structure of the theory correlation matrix, as discussed
in \sec{resum}.

It is straightforward to generalize the above discussion to jet bins with more
jets. For the $N$-jet bin we replace $\sigma_\tot \to \sigma_{\ge N}$,
$\sigma_0\to \sigma_N$, and $\sigma_{\ge 1} \to \sigma_{\ge N+1}$, and take the
appropriate $\sigma_B$. If the perturbative series for $\sigma_{\ge N}$
exhibits large $\alpha_s$ corrections, then the additional large logarithms present in
$\sigma_{\ge N+1}$ will again lead to cancellations when we consider the difference
$\sigma_N = \sigma_{\geq N} - \sigma_{\geq N+1}$. Hence, $\Delta_{\geq N+1}$
will again give a better estimate for the $\Delta_\cut$ that arises from separating
$\sigma_{\geq N}$ into jet bins $\sigma_N$ and $\sigma_{\geq N+1}$.
Another advantage of our procedure is that it is easily generalized to more than two
jet bins by iteration. The case of three jet bins is given in the Appendix.

%%%%%%%%%%%%%%%%%%%%%%%%%%%%%%%%%%%%%%%%%%%%%%%%%%%%%%%%%%%%%%%%%%%%%%%%%%%%%%%%
\section{Example Processes}
\label{sec:examples}
%%%%%%%%%%%%%%%%%%%%%%%%%%%%%%%%%%%%%%%%%%%%%%%%%%%%%%%%%%%%%%%%%%%%%%%%%%%%%%%%

To elucidate the effect of $p_T^\jet$ vetoes on the fixed-order cross sections
and demonstrate our method, we will now go through several explicit examples,
considering in turn $H+0$ jets, $H+1$ jet, $WW + 0$ jets, and $W + 0$, $1$, and
$2$ jets.  All of our NLO $p_T$ spectra are obtained using the \textsc{MCFM}
code~\cite{Campbell:1999ah, Campbell:2002tg, Campbell:2011bn, Campbell:2010cz}.
As our jet algorithm we use anti-$k_T$ for the LHC results and a cone algorithm
for the Tevatron results with $R=0.5$ for both.

%===============================================================================
\subsection{\boldmath Higgs + $0$ Jets}
\label{subsec:H0}
%===============================================================================

\begin{figure*}[t!]
\includegraphics[width=0.495\textwidth]{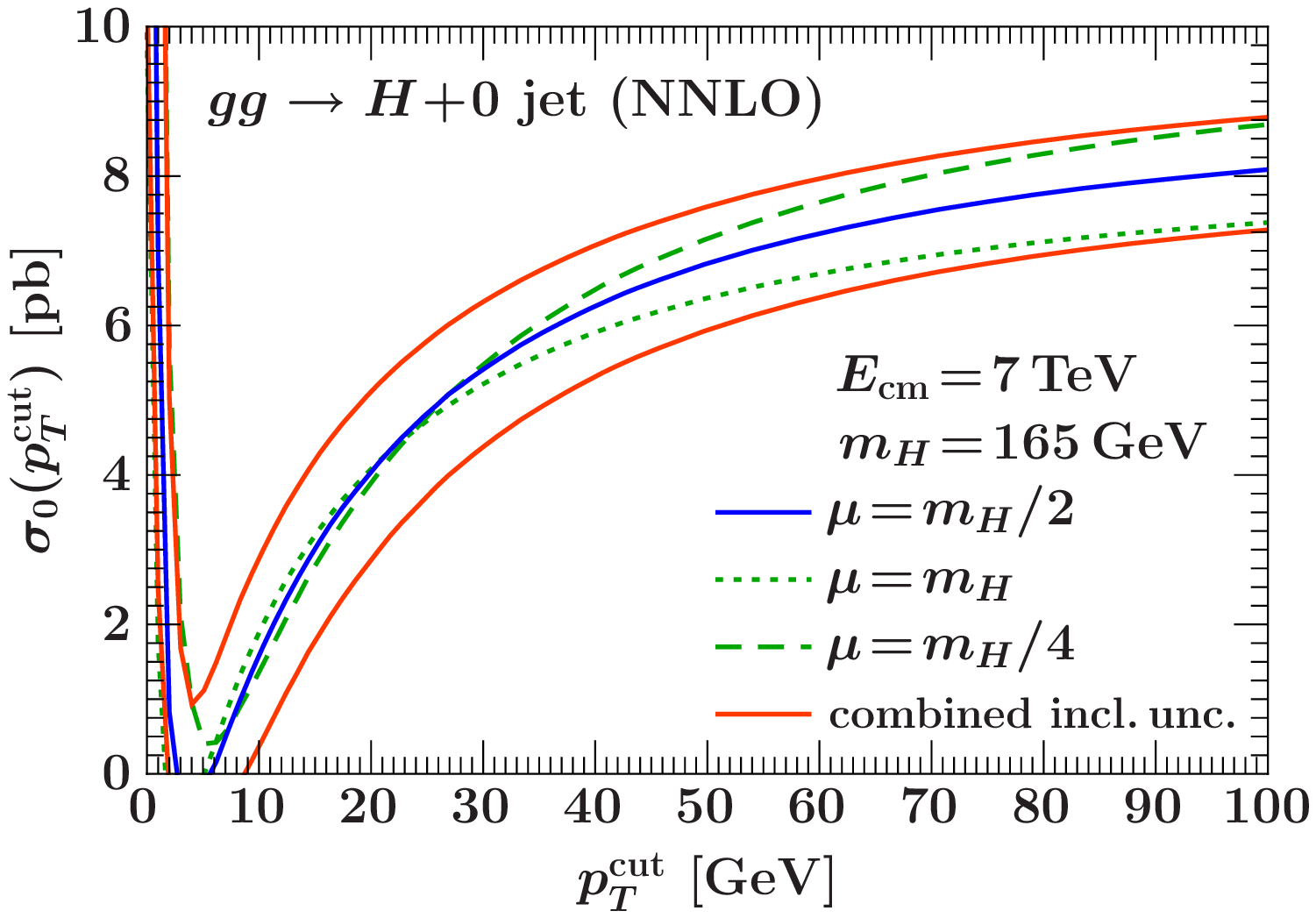}%
\hfill%
\includegraphics[width=0.505\textwidth]{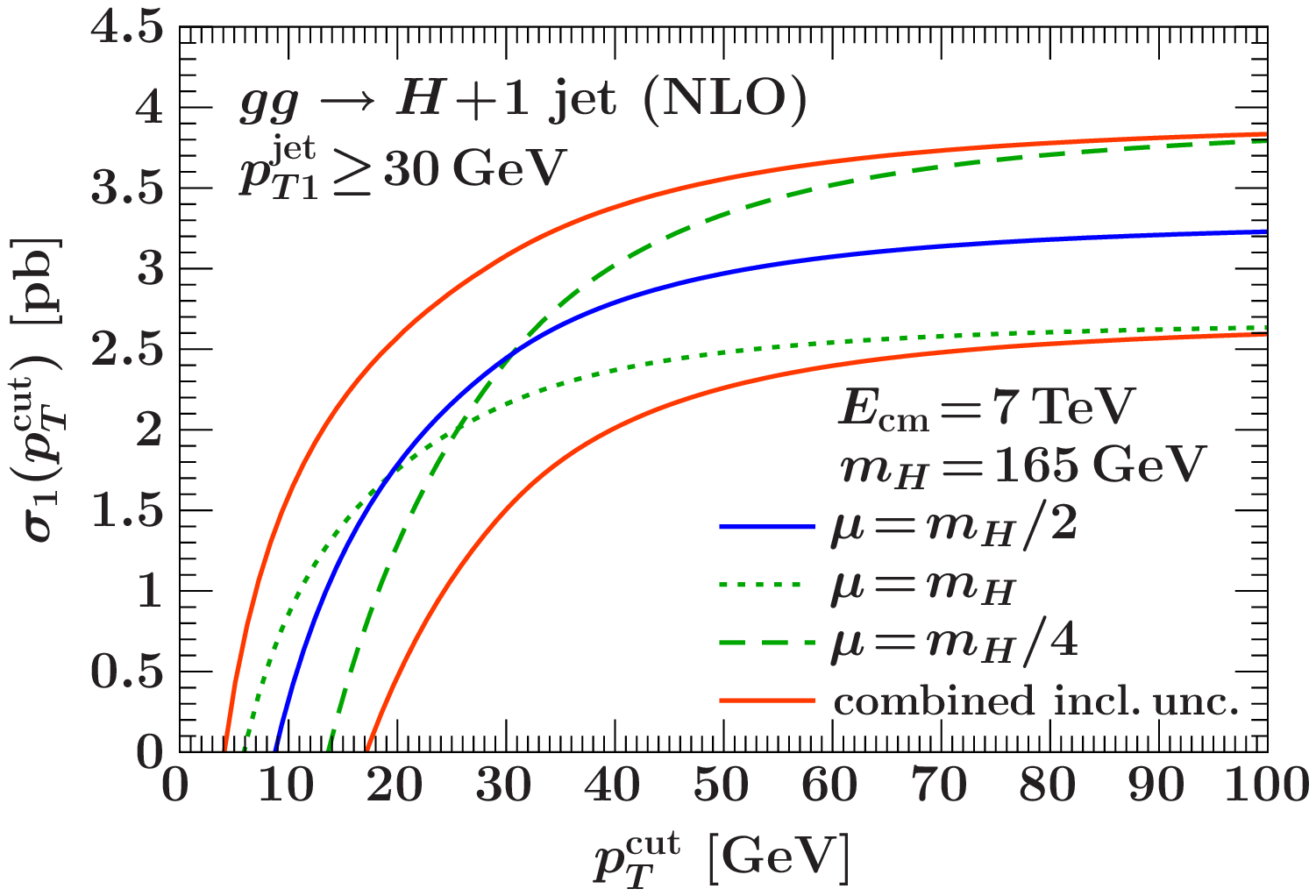}%
\\
\includegraphics[width=0.495\textwidth]{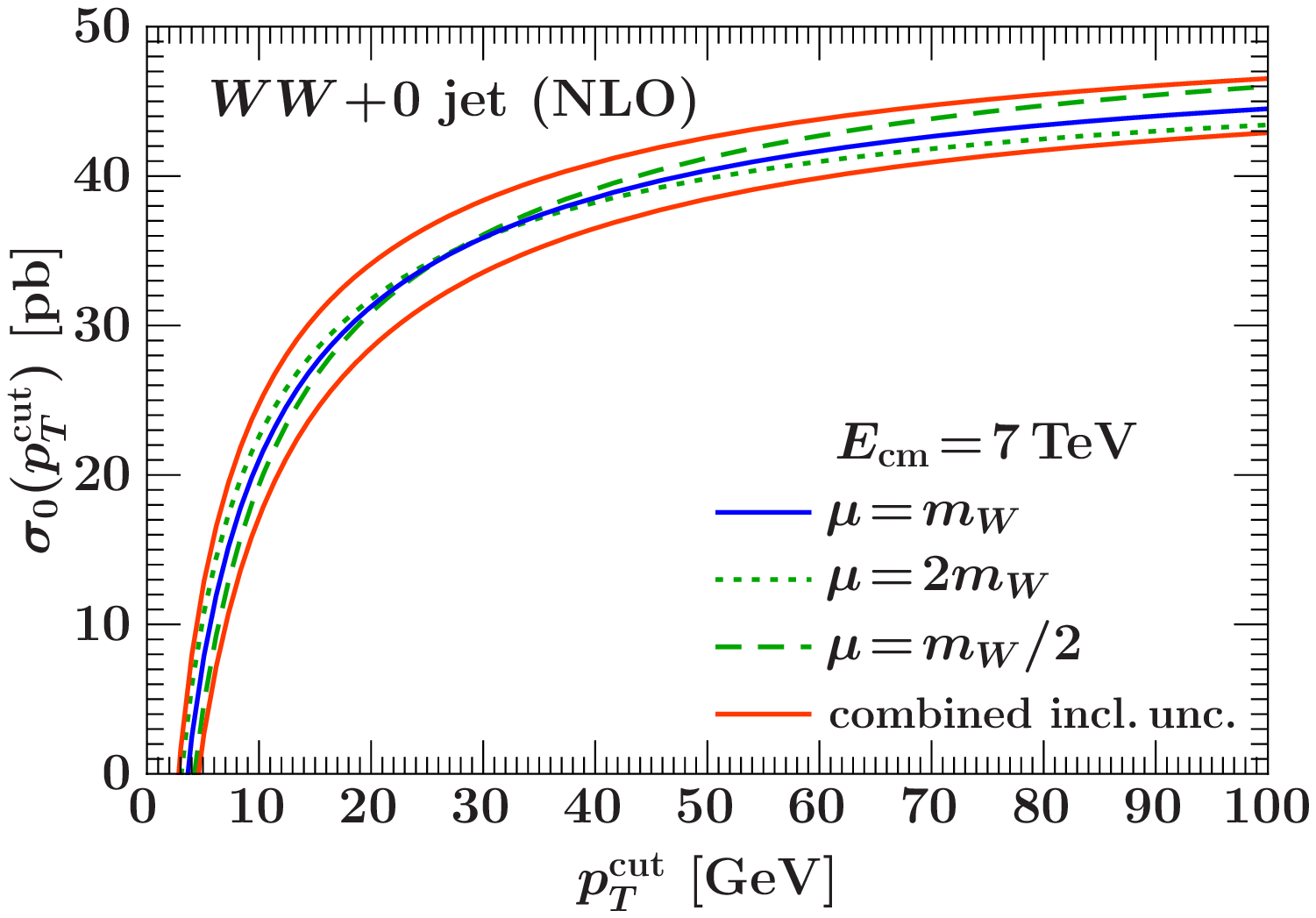}%
\hfill%
\includegraphics[width=0.505\textwidth]{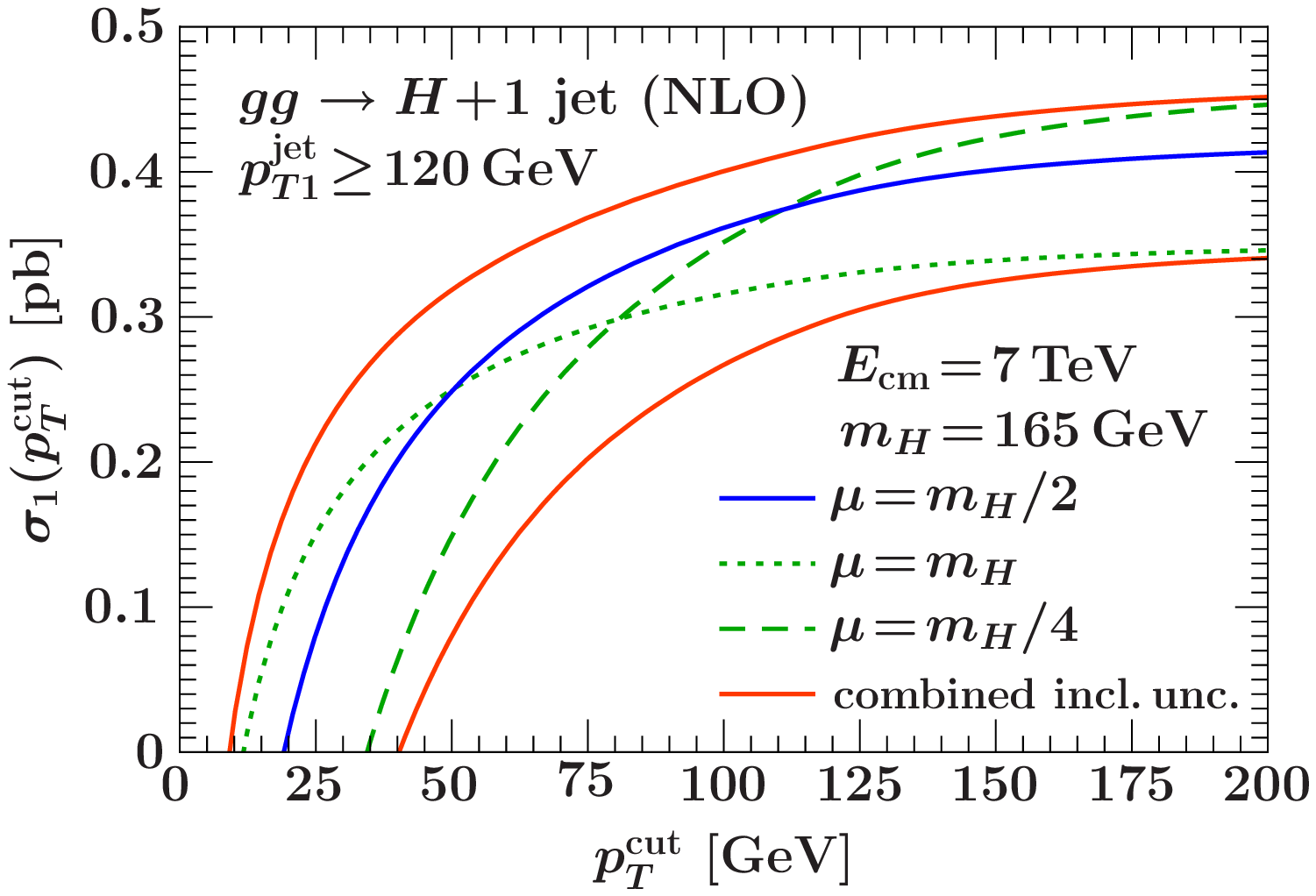}%
\vspace{-0.5ex}
\caption{\label{fig:FOall} Perturbative predictions for $H + 0$ jets (upper left
  panel), $WW + 0$ jets (lower left panel), $H + 1$ jet with $p_{T1}^\jet\geq
  30\GeV$ (upper right panel), and $H + 1$ jet with $p_{T1}^\jet\geq 120 \GeV$
  (lower right panel). Central values are shown by the blue solid curves, direct
  scale variation in the exclusive jet bin by the green dashed and dotted
  curves, and the result of combining independent inclusive uncertainties to get
the jet-bin uncertainty by the outer red solid curves.}
\end{figure*}

\begin{figure*}[t!]
\includegraphics[width=0.5\textwidth]{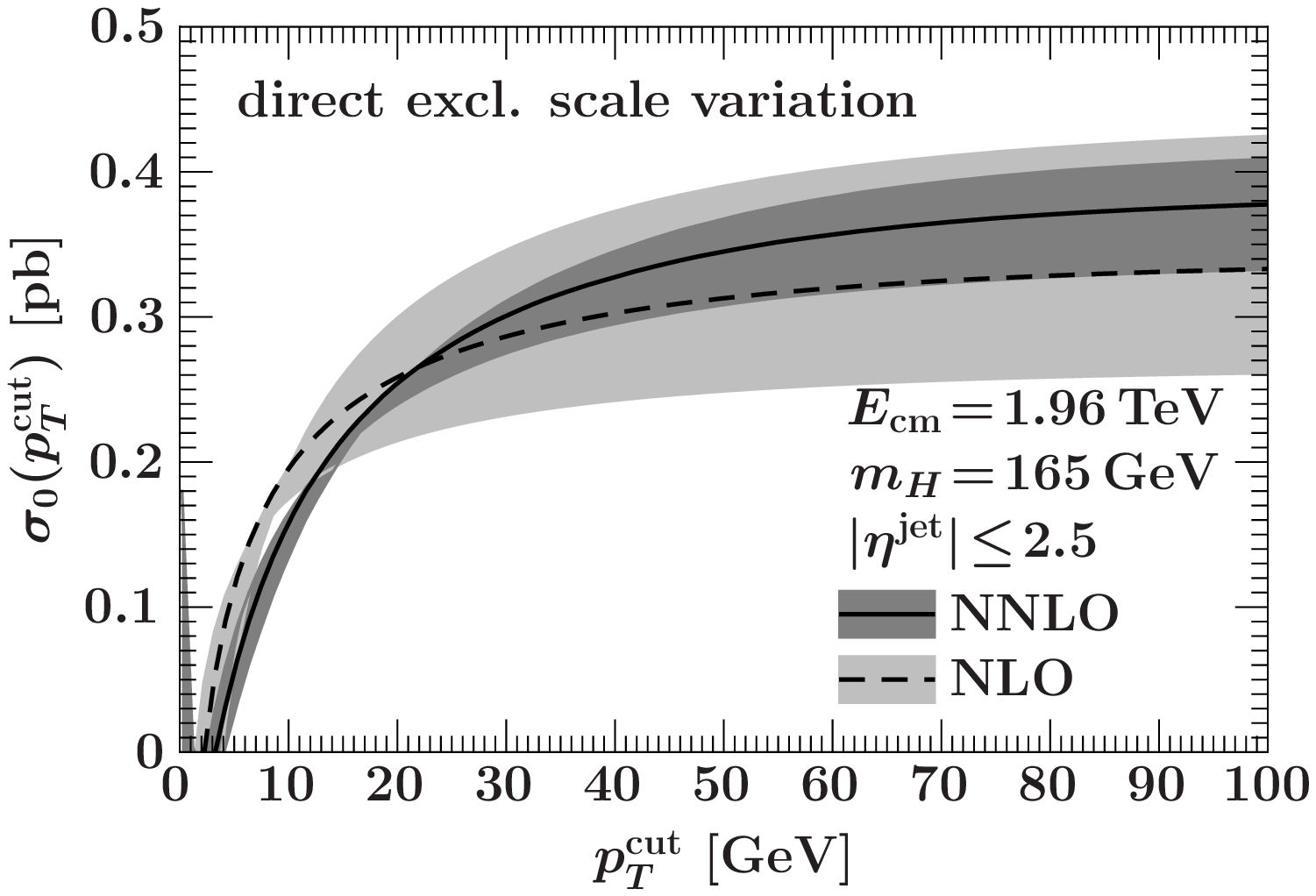}%
\hfill%
\includegraphics[width=0.5\textwidth]{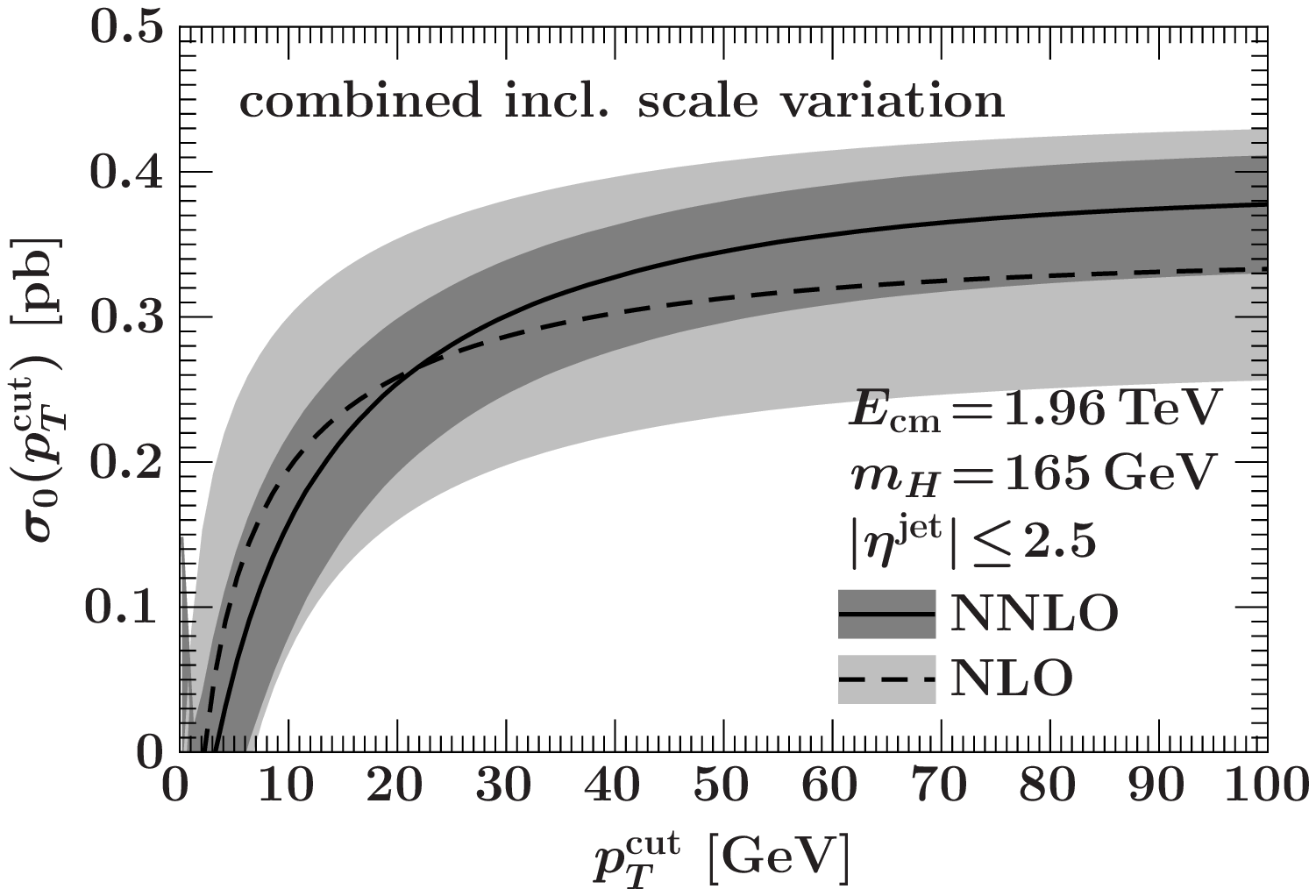}%
\\
\includegraphics[width=0.5\textwidth]{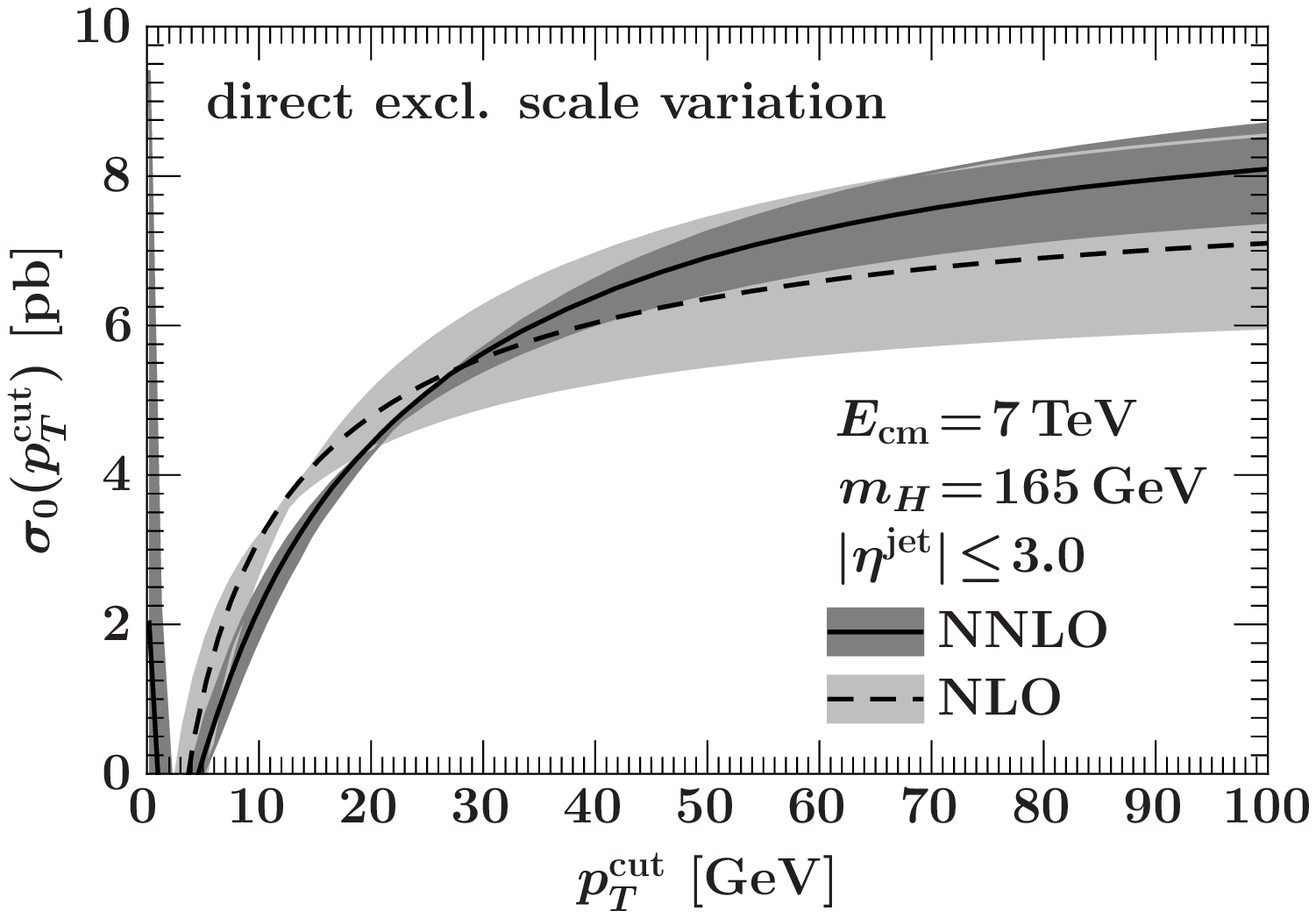}%
\hfill%
\includegraphics[width=0.5\textwidth]{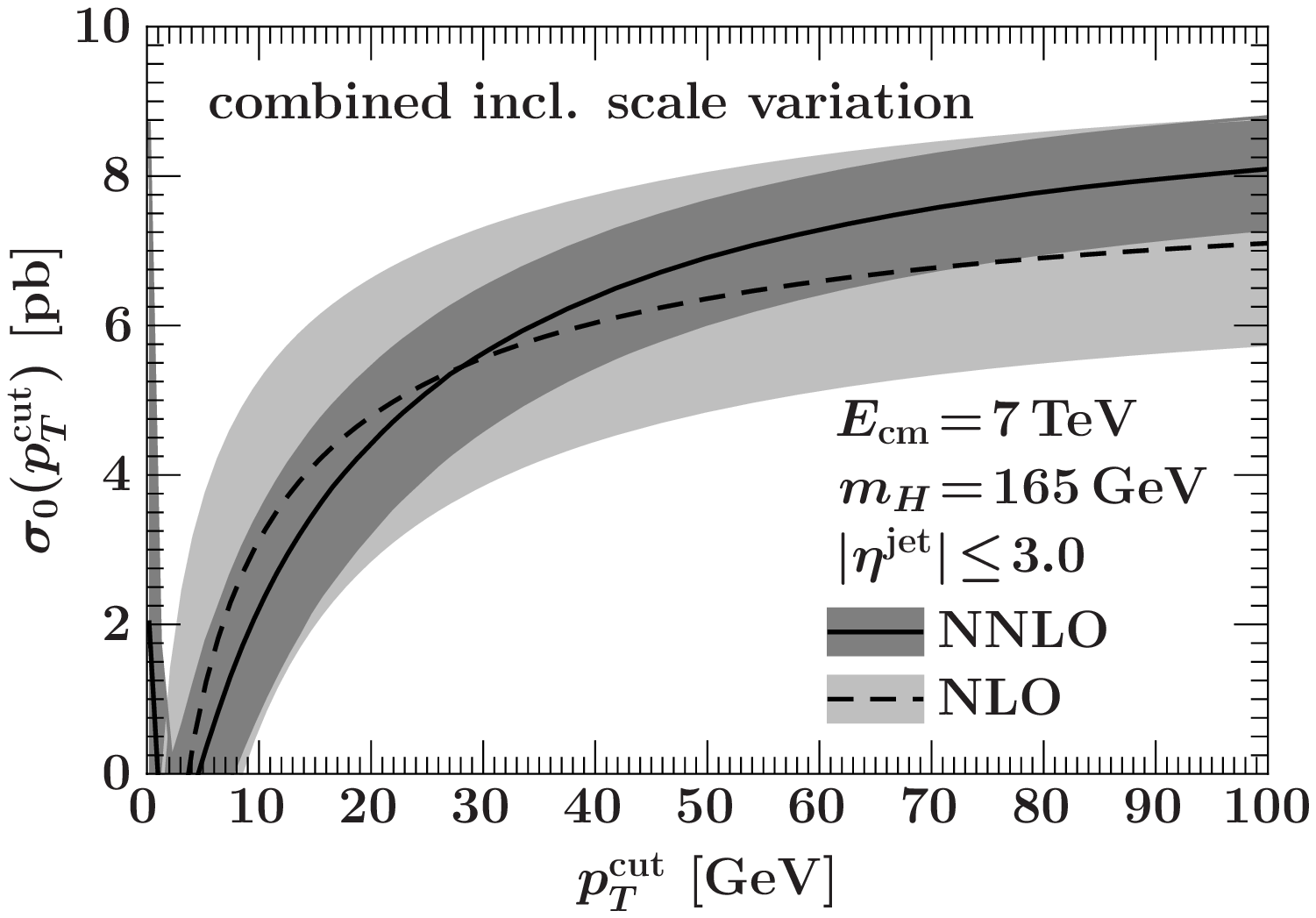}%
\vspace{-0.5ex}
\caption{\label{fig:FOunc} Fixed-order perturbative uncertainties for $gg\to H +
  0$ jets at NLO and NNLO. The upper panels are for the Tevatron and the lower panels
  for the LHC with $\Ecm = 7\TeV$. On the left, the uncertainties are obtained
  from the direct scale variation in $\sigma_0(p_T^\cut)$ between $\mu = m_H/4$
  and $\mu = m_H$. On the right, the uncertainties are obtained by independently
  evaluating the scale uncertainties in $\sigma_\tot$ and $\sigma_{\geq
    1}(p^\cut)$ and combining them in quadrature.  (For the LHC case the dark
  shaded NNLO bands correspond to results in the top left panel of \fig{FOall}.
  The direct exclusive scale variation band corresponds to the dashed green
  lines, and the combined inclusive uncertainty band corresponds to the solid
  red lines.)  }
\end{figure*}

In Higgs production via gluon fusion the cross section is known to
next-to-next-to-leading order (NNLO)~\cite{Dawson:1990zj, Djouadi:1991tka, Spira:1995rr, Harlander:2002wh,
  Anastasiou:2002yz, Ravindran:2003um, Pak:2009dg, Harlander:2009my}, and
exhibits large perturbative corrections.   Consider
the numerical results for the Higgs production cross section for $m_H =
165\GeV$, $\mu_f = \mu_r = m_H/2$, and MSTW2008 NNLO
PDFs~\cite{Martin:2009iq}, for which $\alpha_s \equiv \alpha_s(m_H/2) = 0.1189$.
Here one finds~\cite{Anastasiou:2004xq, Anastasiou:2005qj, Catani:2007vq, Grazzini:2008tf}
%%%
\begin{equation} \label{eq:sigmatotLHC}
\sigma_\tot = (3.32 \pb) \bigl[1 + 9.5\,\alpha_s + 35\,\alpha_s^2 + \ord{\alpha_s^3} \bigr]
\,,\end{equation}
%%%
for the LHC at $\Ecm = 7\TeV$. Note that there
is an $\alpha_s^2$ in the Born cross section, $\sigma_B=3.32\pb$, but only the
relative size of the corrections is important for our discussion. For the Tevatron
the series is
%%%
\begin{equation} \label{eq:sigmatotTev}
\sigma_\tot = (0.15 \pb) \bigl[1 + 9.0\,\alpha_s + 34\,\alpha_s^2 + \ord{\alpha_s^3} \bigr]
\,.\end{equation}
%%%
In both cases the large $K$ factors are clearly visible.%
\footnote{Using instead $\mu_f = \mu_r = m_H$ the coefficients of the $\alpha_s$
and $\alpha_s^2$ terms increase to $11$ and $65$ for the LHC and $12$ and $74$ for the Tevatron, respectively.
The $\alpha_s$ coefficients for the Tevatron for example arise as
$9.0 = 4.9 + 2.0 + 2.1$ ($\mu = m_H/2$) and $12.0 = 4.9 + 5.7 + 1.4$ ($\mu = m_H$) where the three contributions are respectively from the terms in the partonic cross section proportional to $\delta(1-z)$, terms involving the plus functions $[1/(1-z)]_+$ and $[\ln(1-z)/(1-z)]_+$, and the remaining terms that are nonsingular for $z\to 1$. When separating these different terms we keep the overall $1/z$ factor in the convolution integral with measure $\df z/z$.}
For the inclusive $1$-jet cross section at the LHC one finds
%%%
\begin{align} \label{eq:sigma1LHC}
&\sigma_{\geq 1}\bigl(p_T^\jet \geq 30\GeV, \abs{\eta^\jet} \leq 3.0\bigr)
\nn\\* &\qquad
= (3.32 \pb) \bigl[4.7\,\alpha_s + 26\,\alpha_s^2 + \ord{\alpha_s^3} \bigr]
\,,\nn\\*
&\sigma_{\geq 1}\bigl(p_T^\jet \geq 25\GeV)
\nn\\* &\qquad
= (3.32 \pb) \bigl[6.0\,\alpha_s + 32\,\alpha_s^2 + \ord{\alpha_s^3} \bigr]
\,.\end{align}
%%%
The first values correspond to the ATLAS and CMS reference cuts, and the second
to current ATLAS and CMS $H\to WW^*$
analyses~\cite{Chatrchyan:2011tz,Aad:2011kk}. Similarly, for the
typical cuts used in $H\to WW^*$ at the Tevatron~\cite{Aaltonen:2011gs}, one
finds
%%%
\begin{align} \label{eq:sigma1Tev}
&\sigma_{\geq 1}\bigl(p_T^\jet \geq 20\GeV, \abs{\eta^\jet} \leq 2.5\bigr)
\nn\\ &\qquad
= (0.15\pb) \bigl[4.1\,\alpha_s + 27\,\alpha_s^2 + \ord{\alpha_s^3} \bigr]
\,.\end{align}
%%%
In both \eqs{sigma1LHC}{sigma1Tev} one clearly sees the impact of the large
logarithms on the perturbative series. Comparing to
\eqs{sigmatotLHC}{sigmatotTev} one also sees the sizable numerical cancellation
between the two series at each order in $\alpha_s$. The extent of this
cancellation depends sensitively on the value of $p^\cut$.

The perturbative uncertainties on these inclusive cross sections can now be used
to determine the exclusive cross section uncertainties. Varying the scale up and
down by a factor of $2$ around $m_H/2$ gives for the Tevatron
$\sigma_\tot = (0.386 \pm 0.040) \pb$ and $\sigma_{\geq 1} = (0.132 \pm 0.034) \pb$
with the $p_T^\jet$ and $\eta^\jet$ cuts as in \eq{sigma1Tev}.
Adding these in quadrature according to the upper-left entry in \eq{01matrix} gives
%%%
\begin{equation}
  \sigma_0 = (0.254 \pm  0.052) \pb
\,,\end{equation}
%%%
i.e., a $20\%$ uncertainty.  In contrast, when doing a scale variation
directly in the fixed-order expansion for $\sigma_0(p^\cut)$, as in \eq{Tevunc},
one implicitly assumes that the perturbative uncertainties between the series
for $\sigma_\tot$ and $\sigma_{\ge 1}$ are 100\% correlated, giving
$\sigma_0=(0.254\pm 0.006)$. Here this leads to an underestimate for the
remaining uncertainty.  For the LHC, using the reference cuts, we get
$\sigma_\tot = (8.70 \pm 0.75) \pb$ and $\sigma_{\geq 1} = (3.08 \pm 0.59) \pb$,
leading to
%%%
\begin{equation}
  \sigma_0 = (5.63 \pm  0.96) \pb
\,,\end{equation}
%%%
i.e., a $17\%$ uncertainty. In contrast, the direct scale variation for
$\sigma_0$ yields $\sigma_0=(5.63 \pm 0.15)$, which is again an underestimate.

The two procedures of evaluating uncertainties can be compared as a function of
$p_T^\cut$, and in the upper left panel of \fig{FOall} we do so for
$\sigma_0(p_T^\cut)$ for Higgs production. Results for $\sigma_0$ are obtained
at NNLO for the LHC at $\Ecm = 7\TeV$, using \textsc{MCFM} to calculate the $p_T^\cut$
dependence, \textsc{FEHiP}~\cite{Anastasiou:2004xq, Anastasiou:2005qj} for the total NNLO
cross section, and $\mu=m_H/2$ for central values. The central value is the
solid blue curve, and the green dashed and dotted lines show the results of
direct exclusive scale variation by a factor of $2$. For small values of
$p_T^\cut$ the cancellations that take place for $\sigma_0(p^\cut)$ cause the
error bands to shrink. In particular, the direct exclusive scale uncertainty
vanishes at $p_T^\cut\simeq 25\GeV$, where there is an almost exact cancellation
between the two series in \eq{sigma0}, and the uncertainty curves pinch
together. In contrast, the outer red solid lines show the result of our method,
which combines the independent inclusive uncertainties to obtain the exclusive
uncertainty, $\Delta_0^2 =\Delta_\tot^2 + \Delta_{\ge 1}^2$. One can see that
for large values of $p_T^\cut$ this combined inclusive uncertainty estimate
reproduces the direct exclusive scale variation, since $\sigma_{\geq 1}(p^\cut)$
becomes small. On the other hand, for small values of $p_T^\cut$ the
uncertainties obtained in this way are now more realistic, because they
explicitly take into account the large logarithmic corrections.  The features of
this plot are quite generic. In particular, the same pattern of uncertainties is
observed for the Tevatron, when we take $\mu=m_H$ as our central curve with
$\mu=2m_H$ and $\mu=m_H/2$ for the range of scale variation, and whether or not
we only look at jets at central rapidities. We also note that using independent
variations for $\mu_f$ and $\mu_r$ does not change this picture, in particular
the $\mu_f$ variation for fixed $\mu_r$ is quite small.

Since both NLO and NNLO results for $\sigma_0(p_T^\cut)$ are available, it is
also useful to consider the convergence, which we show in \fig{FOunc} for the
Tevatron (top panels) and the LHC at $7\TeV$ (bottom panels). In the left panels we
directly vary the scales in
$\sigma_0(p_T^\cut)$ to estimate the uncertainty, while in the right panels we
again propagate the uncertainties from the inclusive cross sections. As we lower
$p_T^\cut$, the direct exclusive scale variation uncertainty estimate decreases
at both NLO and NNLO, and eventually becomes very small when the curves pinch
and the uncertainty is clearly underestimated. In contrast, the combined inclusive
scale variation gives realistic uncertainties for all values of $p_T^\cut$. In
particular, there is considerable uncertainty for small $p_T^\cut$ where the
summation of logarithms is important.

%===============================================================================
\subsection{\boldmath Higgs + $1$ Jet}
\label{subsec:H1}
%===============================================================================

As our next example we consider the $1$-jet bin in Higgs production from gluon
fusion. This jet bin is defined by two cuts, one which ensures that the
jet with the largest $p_T$ is outside the $0$-jet bin, $p_{T1}^\jet\ge
p_{T1}^\cut$, and one which ensures that the jet with the next largest $p_T$ is
restricted, $p_{T2}^\jet \le p_T^\cut$, so that we do not have two or more
jets. The $1$-jet cross section can be computed as a difference of inclusive
cross sections with these cuts,
%%%
\begin{equation} \label{eq:sig1H}
 \sigma_1 = \sigma_{\ge 1}\bigl(p_{T1}^\jet\ge p_{T1}^\cut\bigr)
 - \sigma_{\ge 2}\bigl(p_{T1}^\jet\ge p_{T1}^\cut,p_{T2}^\jet \ge p_T^\cut\bigr) \,.
\end{equation}
%%%
For convenience we adopt the notation that $p_T^\cut$ is always used for the
cutoff that determines the upper boundary of the jet bin under consideration,
which gives the analog of the $L$ dependent terms in \eq{sigma0}.

The inclusive cross section $\sigma_{\ge 1}$ that includes the $1$-jet bin
exhibits large perturbative corrections, much as $\sigma_\tot$ does for the $0$-jet
bin. For $\sigma_{\ge 1}$ the large corrections are caused in part by the
large double logarithmic series in $\ln(p_{T1}^\jet/m_H)$, but remains
predominantly independent of the large double logarithms of $L=\ln(p_{T2}^\jet/m_H)$
which control the series for $\sigma_{\ge 2}$. With $\mu_f=\mu_r=m_H/2$,
$m_H=165\GeV$, and MSTW2008 NNLO PDFs, we find
%%%
\begin{align} \label{eq:sigma1J1LHC}
&\sigma_{\geq 1}\bigl(p_{T1}^\jet \geq 30\GeV)
\nn\\* &\qquad
= (2.00 \pb) \bigl[1 + 5.4\,\alpha_s + \ord{\alpha_s^2} \bigr]
\,, \nn\\
& \sigma_{\geq 2}\bigl(p_{T1}^\jet\geq 30\GeV, p_{T2}^\jet \geq 30\GeV)
\nn\\ &\qquad
= (2.00 \pb) \bigl[3.6\,\alpha_s +\ord{\alpha_s^2} \bigr]
\,.\end{align}
%%%
For $\sigma_1=\sigma_{\ge 1}-\sigma_{\ge 2}$ there is a sizable cancellation
between these $\alpha_s$ terms. If we lower the cut to $p_{T2}^\jet \geq 22\GeV$
then the logarithm increases and there is an almost exact cancellation with the
$5.4\alpha_s$. In the top right panel of \fig{FOall} we plot $\sigma_1$ as a
function of $p_T^\cut$, and we again see that this cancellation occurs in a
region where there is a dramatic decrease in the direct exclusive scale
dependence (green dashed and dotted curves). Using the inclusive uncertainties
for $\sigma_{\ge 1}$ and $\sigma_{\ge 2}$, and adding them in quadrature, gives
the outer solid red curves, which again avoids this problem and provides a more
realistic estimate for the perturbative uncertainty.

Using the result from the Appendix we can examine the full uncertainties and
correlation matrix with $0$, $1$, and $(\ge 2)$-jet bins in Higgs production.  For
the cuts in \eq{sigma1J1LHC} varying the scale by factors of $2$, we have
$\sigma_\tot = (8.70 \pm 0.75)\pb$, $\sigma_{\geq1} = (3.29\pm0.62)\pb$, and
$\sigma_{\geq 2} = (0.85 \pm 0.49)\pb$, corresponding to relative uncertainties
of $8.6\%$, $18.8\%$, and $57\%$, respectively. We let $\delta(x)$ denote the
relative percent uncertainty of the quantity $x$, and $\rho(x, y)$ the correlation
coefficient between $x$ and $y$. The Appendix yields
%%%
\begin{align} \label{eq:pTcorr}
\delta(\sigma_0) &= 18\%
\,,\qquad &
\delta(\sigma_1) &= 32\%
\,,\nn\\
\rho(\sigma_0, \sigma_\tot) &= 0.77
\,,\qquad &
\rho(\sigma_1, \sigma_{\geq 2}) &= -0.62
\,,\nn\\
\rho(\sigma_0, \sigma_1) &= -0.50
\,,\end{align}
%%%
where we have only shown the nonzero correlations. Note that $\sigma_0$ and
$\sigma_1$ as well as $\sigma_1$ and $\sigma_{\geq 2}$ have a substantial
negative correlation because of the jet-bin boundary they share, while $\sigma_0$ and $\sigma_{\geq 2}$ are uncorrelated.

In contrast, the direct exclusive scale variation results in all the cross
sections being $100\%$ correlated. Because of the cancellations between the
perturbative series, this leads to much smaller (and unrealistic) uncertainties,
with our choice of cuts $\delta(\sigma_0) = 2.3\%$ and $\delta(\sigma_1) =
5.5\%$, which is reflected in the pinching of the dotted and dashed green lines in \fig{FOall}.
(Note that increasing the range of scale variation or separately varying $\mu_r$
and $\mu_f$ does not mitigate this problem.) The analog of \eq{Tevunc} for this
example would be
%%%
\begin{equation}
0.62\times 2.3\% + 0.28\times 5.5\% + 0.10 \times 57\% = 8.6\%
\,.\end{equation}
%%%
When all $\sigma_i$ are $100\%$
correlated, $\sigma_0$ is forced to have a smaller relative uncertainty than
$\sigma_\tot$, as in \eq{Tevunc}, since it has to make up for the much larger
uncertainties in $\sigma_{\geq 2}$.

In addition to the cross sections in each jet bin, we can also consider the
relative jet fractions $f_0 = \sigma_0/\sigma_\tot$ and $\sigma_1/\sigma_\tot$,
which are often used in experimental analyses.  The perturbative theory
uncertainties and correlations for the jet fractions follow by standard error
propagation from those in \eq{pTcorr}. The general expressions are given in
the Appendix, and we find
%%%
\begin{align}
\delta(f_0) &= 13\%
\,,\qquad&
\delta(f_1) &= 33\%
\,,\nn\\
\rho(f_0, \sigma_\tot) &= 0.42
\,,\qquad&
\rho(f_1, \sigma_\tot) &= -0.26
\,,\nn\\
\rho(f_0,f_1) &= -0.80
\,.\end{align}
%%%
Comparing to \eq{pTcorr}, the use of jet fractions with $\sigma_\tot$ in the
denominator yields a nonzero anticorrelation for $\sigma_\tot$ with the $1$-jet
bin, and decreases the correlation for $\sigma_\tot$ with the $0$-jet bin.

It is also interesting to consider the case with $p_{T1}^\jet \geq 120\GeV$,
where the logarithms of $p_{T1}^\jet/m_H$ are not large. The cross section
$\sigma_{\ge 1}$ now has a smaller perturbative correction, but for a region of
cuts on $p_{T2}^\jet$ there are still substantial cancellations in $\sigma_1$.
For instance, for $p_{T2}^\jet\geq 60\GeV$ we have
%%%
\begin{align} \label{eq:sigma1J1LHC2}
&\sigma_{\geq 1}\bigl(p_{T1}^\jet \geq 120\GeV)
\nn\\ &\qquad
= (0.31\pb) \bigl[1 + 2.9\,\alpha_s + \ord{\alpha_s^2} \bigr]
\,,\nn\\
& \sigma_{\geq 2}\bigl(p_{T1}^\jet \geq 120\GeV, p_{T2}^\jet \geq 60\GeV)
\nn\\ &\qquad
= (0.31 \pb) \bigl[3.7 \,\alpha_s +\ord{\alpha_s^2} \bigr]
\,,\end{align}
%%%
and the $\alpha_s$ terms completely cancel around $p_{T2}^\jet \ge 70\GeV$. In
the bottom right panel of \fig{FOall} we plot $\sigma_1$ as a function of
$p_T^\cut$ for this scenario. Once again the combined inclusive uncertainties
(solid red curves) give a better estimate than the direct exclusive scale
uncertainty determined by up/down $\mu$ variation in $\sigma_1$ (green dotted
and dashed curves). It is interesting to notice that the curves dive and a
logarithmic summation in $p_{T2}^\jet$ becomes important earlier now, i.e., at
much larger values for $p_{T2}^\jet$, when the cut on $p_{T1}^\jet$ is raised.
For $p_{T1}^\jet \geq 120\GeV$ and $p_{T2}^\jet \leq 30\GeV$ fixed-order
perturbation theory does not yield a controlled expansion, and the resummation
of the jet-veto logarithms is clearly necessary.

\begin{figure*}[t!]
\includegraphics[width=0.5\textwidth]{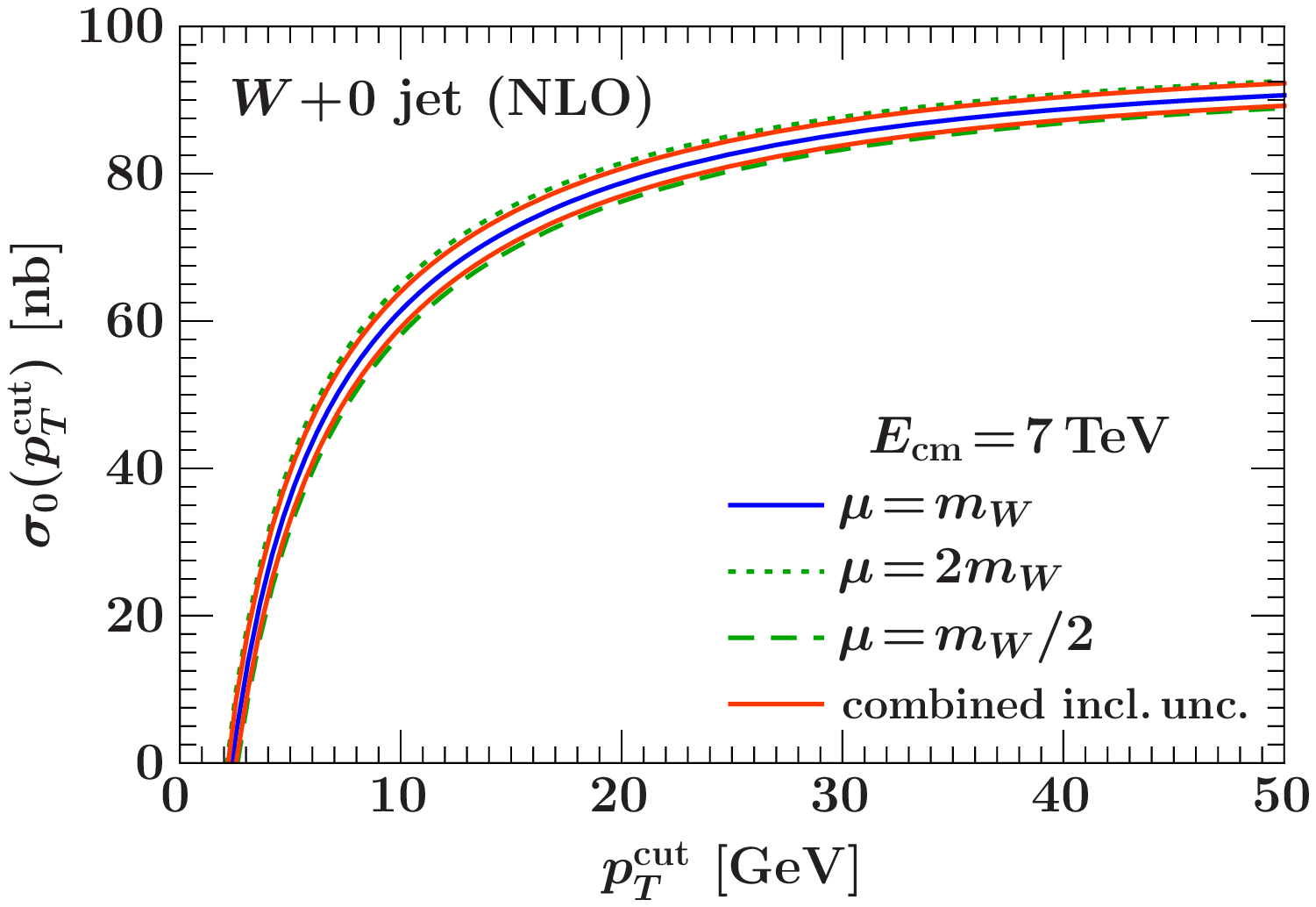}%
\hfill%
\includegraphics[width=0.5\textwidth]{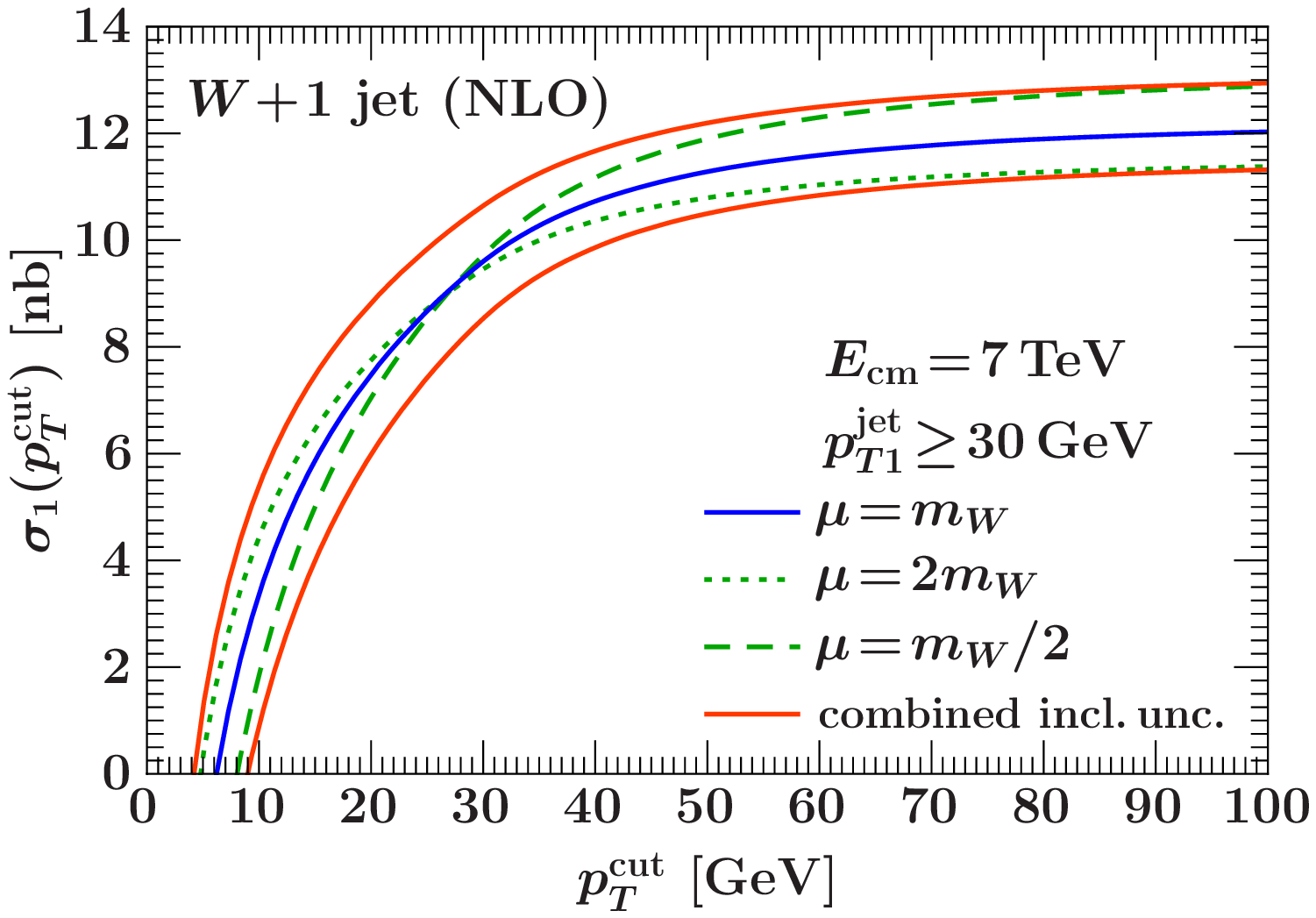}%
\\
\includegraphics[width=0.5\textwidth]{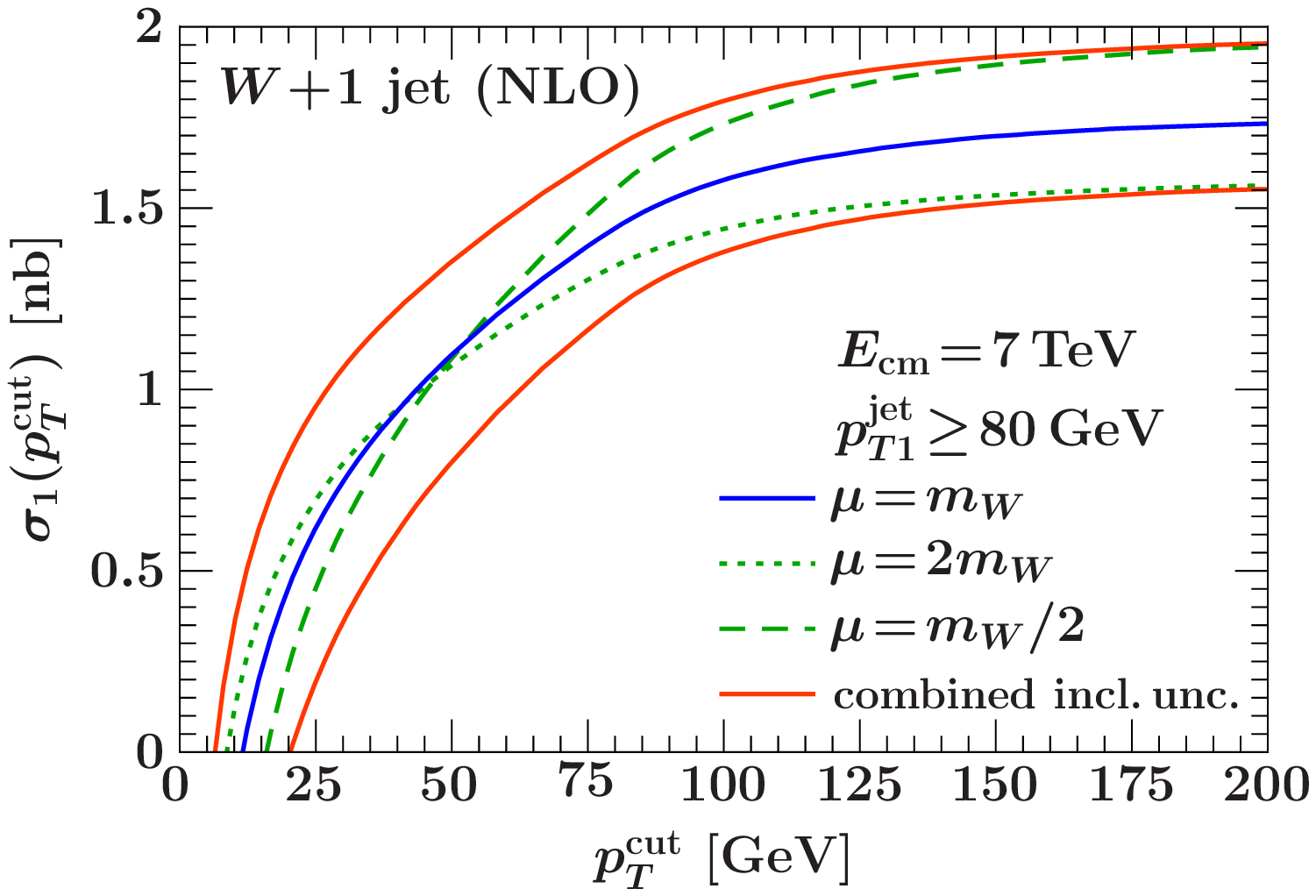}%
\hfill%
\includegraphics[width=0.5\textwidth]{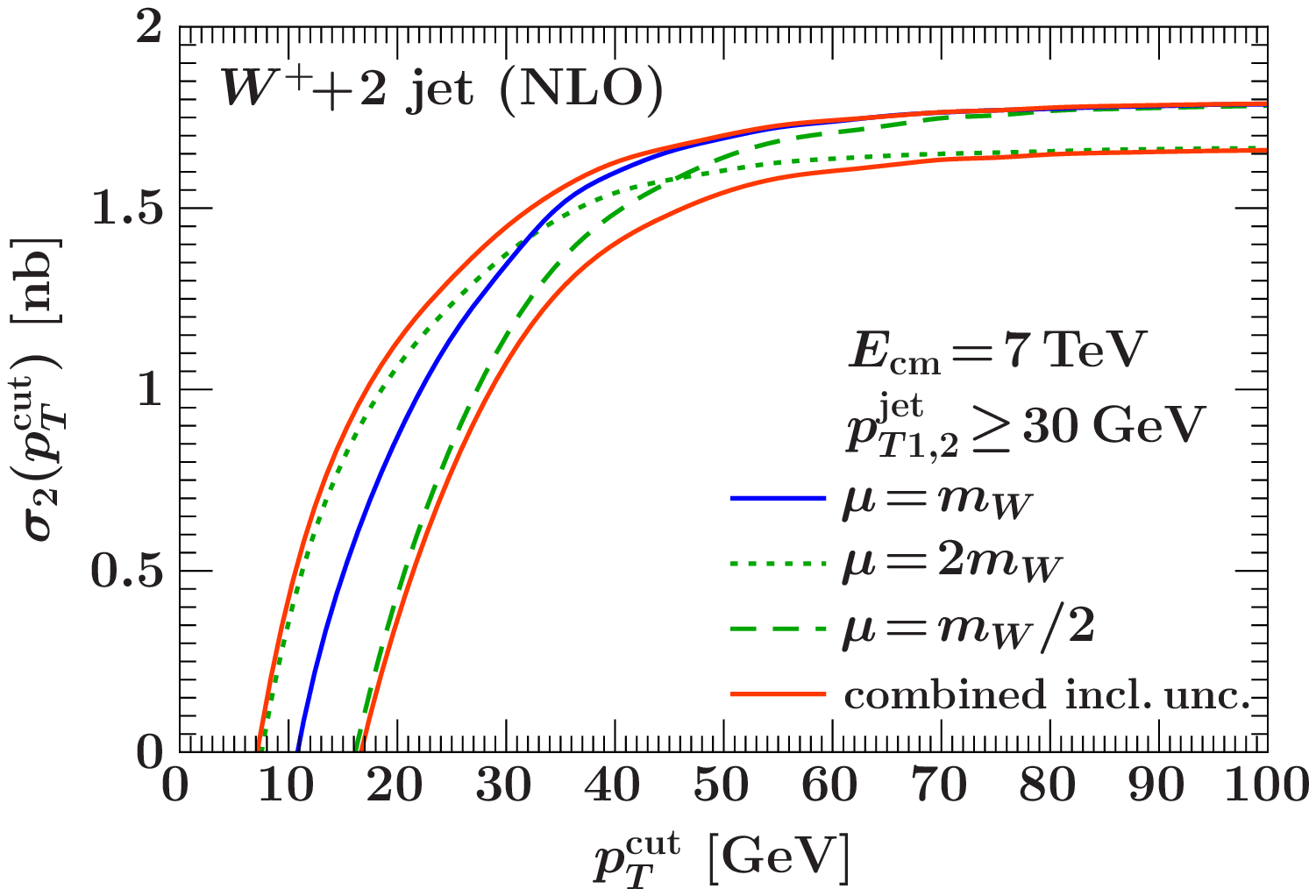}%
\vspace{-0.5ex}
\caption{\label{fig:Wjets} Fixed-order perturbative uncertainties for the
  exclusive $pp\to W + 0, 1, 2$ jet cross sections at NLO for the LHC with $\Ecm
  = 7\TeV$. Central values are shown by blue solid curves, direct exclusive scale variation
  in the exclusive jet bin by the green dashed and dotted curves, and the result
  of combining independent inclusive uncertainties to get the jet-bin
  uncertainty by the outer red solid curves. }
\end{figure*}

%===============================================================================
\subsection{\boldmath $WW + 0$ Jets}
\label{subsec:WW}
%===============================================================================

The process $pp\to WW + 0$ jets is the dominant irreducible background for the
$H\to WW^*$ search in the $0$-jet bin, and also exhibits a relatively large $K$
factor $\sim 1.5$. Hence, it is interesting to contrast the scale uncertainties here
with those found for $H+ 0$ jets. Including the Higgs search cuts
(modulo the jet veto), the $K$ factor for $WW$ becomes larger than
$2$~\cite{Campbell:2011bn}, but we will not include those cuts in our analysis
here. With $\mu_r=\mu_f=m_W$, NLO MSTW2008 PDFs,
and $\alpha_s \equiv \alpha_s(m_W) = 0.1226$, the total $pp\to WW$ cross
section is
%%%
\begin{equation} \label{eq:sigmatotWWLHC}
\sigma_\tot = (32.5 \pb) \bigl[1 + 3.6\,\alpha_s  + \ord{\alpha_s^2} \bigr]
\,,\end{equation}
%%%
while for the inclusive $1$-jet cross section with logarithms of $p_T^\cut$ we have
%%%
\begin{align} \label{eq:sigma1WWLHC}
&\sigma_{\geq 1}\bigl(p_T^\jet \geq 30\GeV)
= (32.5 \pb) \bigl[2.8\,\alpha_s  + \ord{\alpha_s^2} \bigr]
\,.\end{align}
%%%
Thus, when we consider $\sigma_0=\sigma_\tot-\sigma_{\ge 1}$ there is a sizable
cancellation for the $\alpha_s$ terms. In \fig{FOall}, lower left panel, we show
$\sigma_0$ for $pp\to WW + 0$ jets as a function of $p_T^\cut$.  Once again the
dotted and dashed green curves from direct exclusive scale variation exhibit a pinching near $p_T^\cut \sim
30\GeV$ due to cancellations between the two perturbative series in
\eqs{sigmatotWWLHC}{sigma1WWLHC}, leading to an underestimate of the perturbative
uncertainty. The combined inclusive uncertainty estimate again mitigates this problem.
The pattern of uncertainties here is the same as for $H+0$ jets and
$H+1$ jet, just with smaller overall uncertainties. Just like for $H+0$ jets
using independent variations for $\mu_f$ and $\mu_r$ does not change the
picture, the $\mu_f$ variation for fixed $\mu_r$ is again quite small.

%===============================================================================
\subsection{\boldmath $W + 0$ Jets}
\label{subsec:W0}
%===============================================================================

The exclusive process $pp\to W+ N$ jets is an important benchmark
process at the LHC and also an important standard model background for new
physics searches looking for missing energy.
In this section we consider $pp \to W+0$ jets, which provides us with a case to
test our method when the perturbative corrections in the inclusive cross
sections are not as large. For simplicity, we only work to NLO here.
Using $\mu_f = \mu_r = m_W$ for the central value and
MSTW2008 NLO PDFs, the inclusive $W$ production cross section is
%%%
\begin{equation} \label{eq:sigmatotWLHC}
\sigma_\tot = (80.7 \nb) \bigl[1 + 1.3\,\alpha_s  + \ord{\alpha_s^2} \bigr]
\,,\end{equation}
%%%
where we have summed over $W^\pm$, and have not included the leptonic branching
fractions. For the inclusive $1$-jet cross section we have
%%%
\begin{equation} \label{eq:sigma1WLHC}
\sigma_{\geq 1}\bigl(p_T^\jet \geq 30\GeV)
= (80.7 \nb) \bigl[0.9\,\alpha_s  + \ord{\alpha_s^2} \bigr]
\,.\end{equation}
%%%
The perturbative coefficients in \eqs{sigmatotWLHC}{sigma1WLHC} are much smaller
than in Higgs production. The resulting predictions for $\sigma_0(p_T^\cut)$ are
shown in the top left panel of \fig{Wjets}, where the different lines have the
same meaning as in \fig{FOall}. Since the $\alpha_s$ corrections are not very large
here, the $\mu_f$ scale variation in the PDFs dominates over the $\mu_r$ variation
in $\alpha_s$ and produces a $100\%$ negative correlation between $\sigma_\tot$ and
$\sigma_{\geq 1}$. (Keeping $\mu_f$ fixed at $m_W$ and only varying $\mu_r$ results
in the expected pinching of the dotted and dashed green lines.)
This means their scale uncertainties add linearly in $\sigma_0$,
which maximizes the uncertainty in this $0$-jet cross section. In this case,
our method, shown by the outer solid red lines, gives an uncertainty band very similar
to direct exclusive scale variation.  Hence, our method of using independent inclusive
uncertainties still remains consistent for this situation.

%===============================================================================
\subsection{\boldmath $W + 1$ Jet}
\label{subsec:W1}
%===============================================================================

For $pp\to W+1$ jet the perturbative corrections in $\sigma_{\ge 1}$ are larger
than those in the $W$ total cross section, which is in part influenced by
logarithms from the lower cut on $p_{T1}^\jet$, the $p_T$ of the leading jet.
The situation for the $W+1$ jet bin is similar to $H+1$ jet.  Considering
\eq{sig1H} the series for the inclusive $2$-jet cross section, $\sigma_{\ge 2}$,
has large double logarithms $L=\ln(p_{T2}^\jet/m_W)$ of the second largest jet
$p_T$, which are independent of those in the perturbative series for
$\sigma_{\ge 1}$.  Taking $\mu=m_W$ for central values, and using MSTW2008 PDFs
at NLO, the total $W^+ + W^-$ cross sections with both jet cuts at $30\GeV$ are
%%%
\begin{align} \label{eq:sigmaW1JLHC}
&\sigma_{\geq 1}\bigl(p_{T1}^\jet \geq 30\GeV)
\nn\\ &\qquad
= (8.61 \nb) \bigl[1 + 3.4\,\alpha_s + \ord{\alpha_s^2} \bigr]
\,, \nn\\
& \sigma_{\geq 2}\bigl(p_{T1}^\jet\geq 30\GeV, p_{T2}^\jet \geq 30\GeV)
\nn\\ &\qquad
= (8.61 \nb) \bigl[2.5\,\alpha_s +\ord{\alpha_s^2} \bigr]
\,.\end{align}
%%%
Once again the result for $\sigma_{\geq 2}$ and the precise cancellation that
occurs in $\sigma_1$ is quite sensitive to $p_T^\cut$, the cut on $p_{T2}^\jet$,
yielding an almost exact cancellation of the $3.4\,\alpha_s$ for $p_{T2}^\jet
\geq 25\GeV$.
In the top-right panel of \fig{Wjets} we plot $\sigma_1$ as a function of
$p_T^\cut$, with direct exclusive scale variation (green dashed and dotted curves) and
those derived from independent inclusive uncertainties (solid red curves). Just
like for $H+1$ jet, the direct exclusive scale variation curves pinch, while the inclusive
curves avoid this problem and remain realistic.

We can also consider what happens when we make a larger cut on $p_{T1}^\jet$.
Here, unlike for the Higgs case, the relative size of the perturbative correction in
$\sigma_{\ge 1}$ increases. For instance,
%%%
\begin{align} \label{eq:sigmaW1JLHC2}
&\sigma_{\geq 1}\bigl(p_{T1}^\jet \geq 80\GeV)
\nn\\ &\qquad
= (1.07\nb) \bigl[1 + 5.3\,\alpha_s + \ord{\alpha_s^2} \bigr]
\,,\nn\\
& \sigma_{\geq 2}\bigl(p_{T1}^\jet \geq 80\GeV, p_{T2}^\jet \geq 60\GeV)
\nn\\ &\qquad
= (1.07 \nb) \bigl[4.1 \,\alpha_s +\ord{\alpha_s^2} \bigr]
\,.\end{align}
%%%
For $p_{T2}^\jet \ge p_T^\cut$ in $\sigma_{\ge 2}$ the resulting $1$-jet cross
section $\sigma_1$ is shown as a function of $p_T^\cut$ in the bottom-left panel
of \fig{Wjets}. The situation for the uncertainties is similar to that for the
less stringent cut on $p_{T1}^\jet$ in the upper-right panel. Much like in $H+1$
jet the logarithms start to influence the cross section at larger values of
$p_T^\cut$ for the larger $p_{T1}^\jet$ cut.

%===============================================================================
\vspace{-1ex}
\subsection{\boldmath $W + 2$ Jets}
\label{subsec:W2}
%===============================================================================

As our last example we consider $W + 2$ jets, and for simplicity we only
consider the case of $W^+$ production. The inclusive $2$-jet and $3$-jet cross
sections with all jets cut at $30\GeV$ are
%%%
\begin{align} \label{eq:sigmaW2JLHC}
\sigma_{\geq 2}\bigl(p_{T1,2}^\jet \geq 30\GeV)
&= (1.60 \nb) \bigl[1 + 1.0\,\alpha_s + \ord{\alpha_s^2} \bigr]
\,, \nn\\
\sigma_{\geq 3}\bigl(p_{T1,2,3}^\jet\geq 30\GeV)
&= (1.60 \nb) \bigl[2.3\,\alpha_s +\ord{\alpha_s^2} \bigr]
\,,\end{align}
%%%
and the resulting exclusive $2$-jet cross section as a function of the $p_T^\cut$
on the third jet is shown in the bottom-right panel in \fig{Wjets}.

There are two different types of diagrams contributing to this process, those
having two external quark lines and two gluon lines at lowest order ($qqgg$),
and those having four external quark lines at lowest order ($qqqq$). The
$qqgg$-type contributions have the same behavior as $W+1$ jet, again displaying
a pinching in the direct exclusive scale variation curves. On the other hand, in the
$qqqq$-type contributions the PDF scale dependence dominates, similar to what we
observed for $W+0$ jets. The combination of the two leads to the behavior seen
in \fig{Wjets} at large $p_T^\cut$, where the scale uncertainties in the
inclusive $2$-jet cross section are asymmetric. Here there is some choice for
how to combine the scale variation into an uncertainty estimate for
$\sigma_{\geq 2}$ (green dashed and dotted curves).  The choice one makes for
$\sigma_{\geq 2}$ simply propagates into the equivalent choice for the exclusive
$2$-jet bin $\sigma_2$ (solid red curves).  For simplicity in the bottom-right
panel of \fig{Wjets} we still use $\mu = m_W/2$ and $\mu = 2m_W$ to determine
$\Delta_{\geq 2}$, in which case the central value should be taken as the center
of the band rather than the dark solid blue line for $\mu = m_W$.

For $W+2$ jets in \fig{Wjets} the pinching caused by the $qqgg$
contributions is again mitigated by combining the inclusive uncertainties.
Hence, we see that our method can be applied and gives more stable uncertainty
estimates even in more complicated cases where several components contribute to
the cross section.

Note that we have also checked that when increasing the cuts on the two leading
jets, the same effect as in $H+1$ jets and $W+1$ jets happens here as well.
Namely, the jet-veto logarithms from restricting the third jet become more
important earlier and influence the cross section at larger values of $p_T^\cut$
for larger $p_{T1,2}^\jet$ cuts.

%%%%%%%%%%%%%%%%%%%%%%%%%%%%%%%%%%%%%%%%%%%%%%%%%%%%%%%%%%%%%%%%%%%%%%%%%%%%%%%%
\section{\boldmath Resummation for Higgs + $0$ jets}
\label{sec:resum}
%%%%%%%%%%%%%%%%%%%%%%%%%%%%%%%%%%%%%%%%%%%%%%%%%%%%%%%%%%%%%%%%%%%%%%%%%%%%%%%%

\begin{figure*}[ht!]
\includegraphics[width=0.505\textwidth]{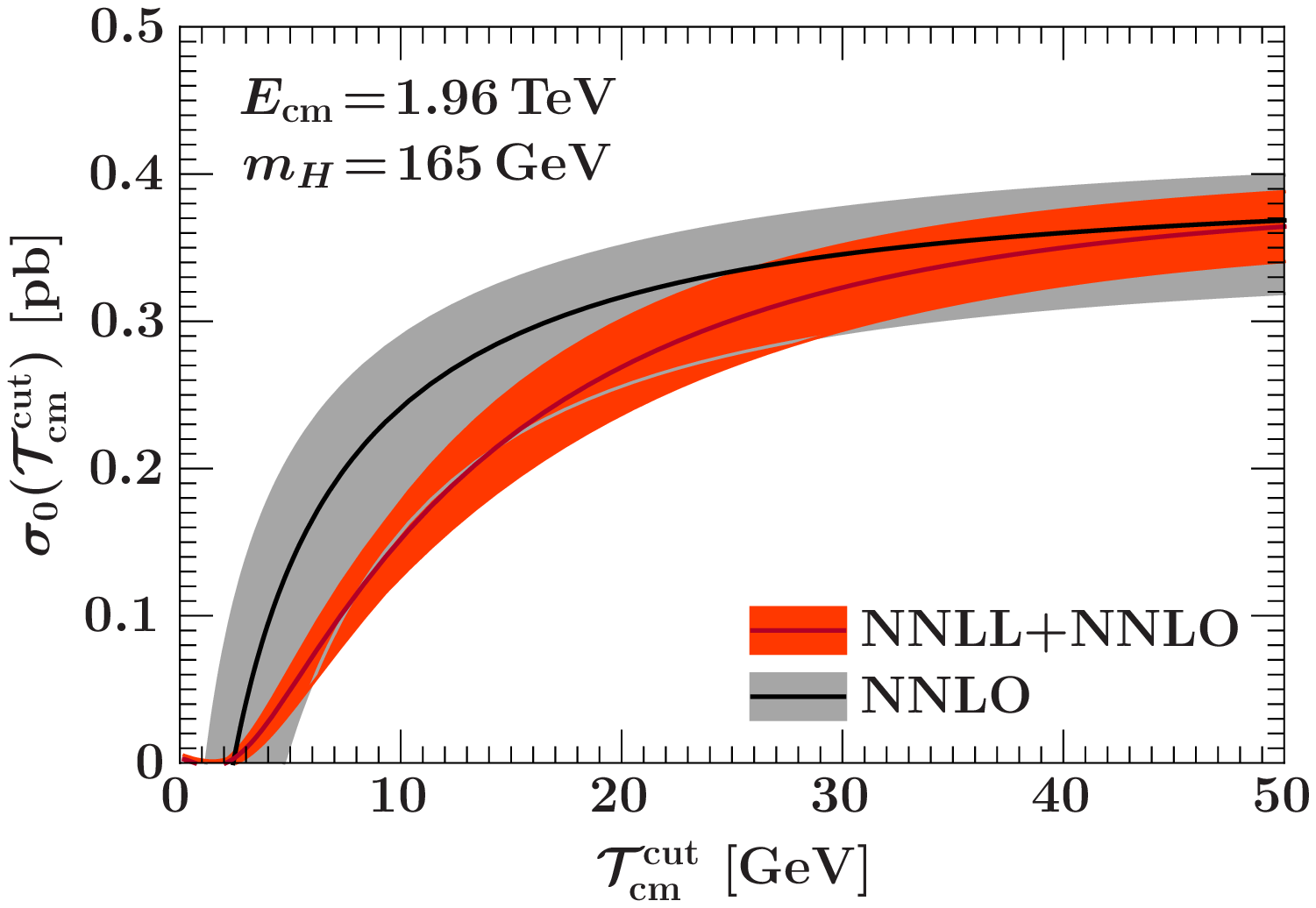}%
\hfill%
\includegraphics[width=0.495\textwidth]{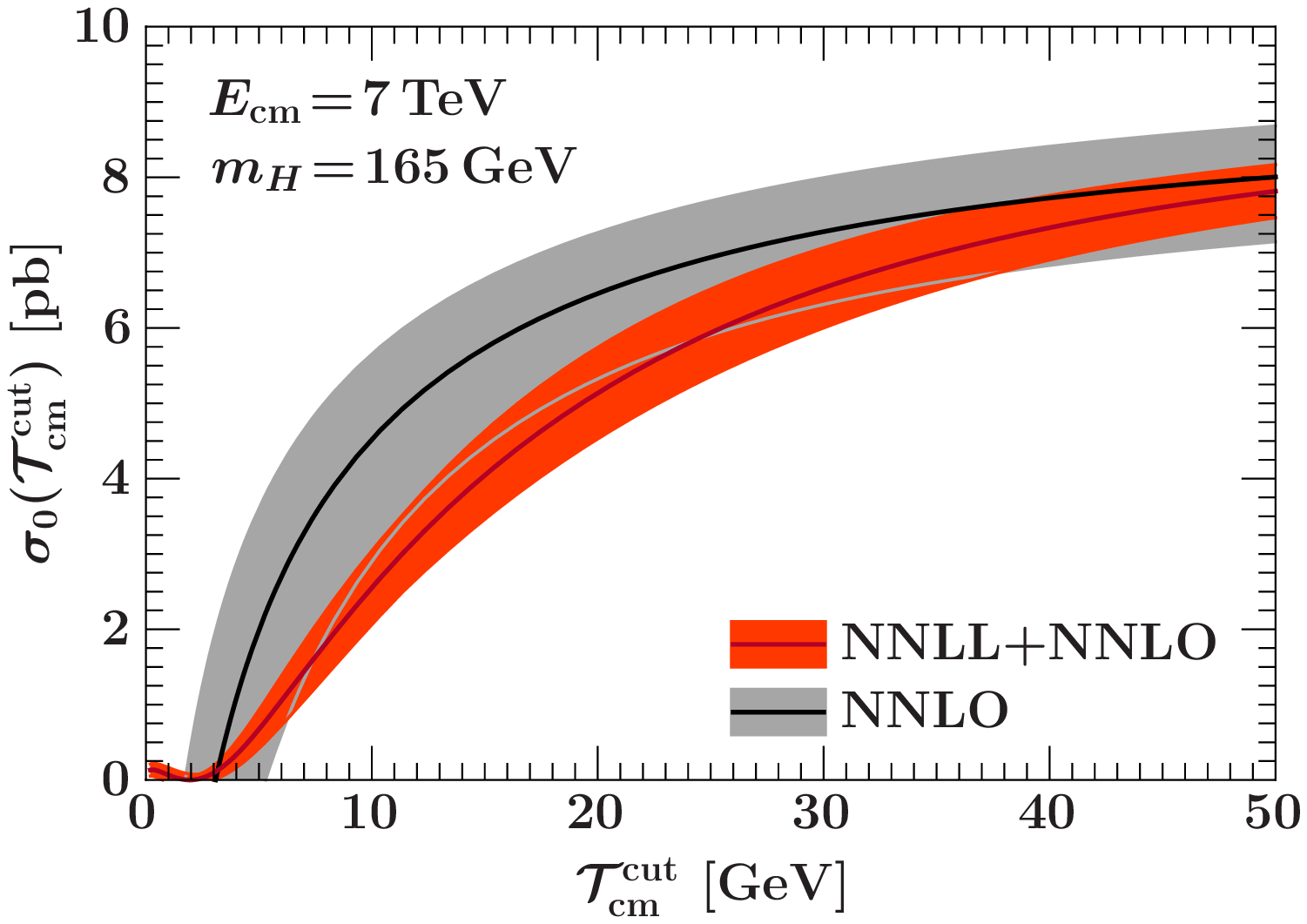}%
\vspace{-0.5ex}
\caption{\label{fig:resum} Comparison of $gg\to H + 0$ jets using $\Tcmc$ at
  fixed NNLO with the resummed results at NNLL+NNLO. For the fixed-order
  uncertainties in $\sigma_0$ we use independent inclusive scale variations in
  $\sigma_\tot$ and $\sigma_{\ge 1}$.  The uncertainty method for the resummed
  results is described in the text.}
\end{figure*}

In \sec{examples} we have seen that direct exclusive scale variation often leads to an
accidental underestimate of the uncertainties for exclusive jet bin cross sections for a
range of experimentally relevant cuts. Instead combining independent
uncertainties on inclusive cross sections yields a more uniform (and larger)
uncertainty band for the exclusive jet bins. The region where direct exclusive scale
variation runs into trouble borders the region where the resummation of the
large logarithms of $p_T^\jet$ becomes important.  In this section, we test how
realistic the fixed-order scale uncertainties are by comparing them to a case
where the resummation of large logarithms induced by the jet bin are known to
NNLL+NNLO accuracy.

We again consider $H+0$ jets from gluon fusion. At NNLL order accuracy the
resummation is sensitive to the precise jet algorithm used to define $p_T^\jet$,
and other complications in the required theoretical setup.  To avoid these
issues, we will use a slightly different variable to define the $0$-jet bin, an
inclusive event shape known as beam thrust~\cite{Stewart:2009yx},
%%%
\begin{equation} \label{eq:Tau_def}
\Tcm = \sum_k (E_k - |p_k^z|)
\,.\end{equation}
%%%
The sum over $k$ runs over all particles except the Higgs decay products.  Beam
thrust essentially measures the thrust of an event along the $\hat z$ beam axis.
When $\Tcm \leq \Tcmc$, from \eq{Tau_def} we see that events in $\sigma_0(\Tcm)$
are only allowed to contain hard radiation in the forward regions at large
rapidities, and hence this cut vetoes central jets. Much like with $p_T^\jet$
the perturbative series for this $\sigma_0$ has double logarithms, for example
the analog of \eq{sig0dbleL} is
%%%
\begin{align}
\sigma_0(\Tcmc) &= \sigma_B \Bigl(1 - \frac{3\alpha_s}{\pi}\, \ln^2
  \frac{\Tcmc}{m_H} + \dotsb \Bigr)
\,.\end{align}
%%%

For beam thrust, the all-order resummation of perturbative corrections is known
to NNLL order for both $H+0$ jets and $V+ 0$ jets~\cite{Stewart:2010qs,
  Stewart:2010pd, Berger:2010xi}.  For Higgs production the computation has been
extended to fully include all NNLO corrections, and it was observed that the
resummed cross section at NNLL+NNLO had larger uncertainties than the pure NNLO
result for $\sigma_0(\Tcm\leq \Tcmc)$ utilizing direct exclusive fixed-order
scale variation. This led to the conclusion that the direct exclusive scale
variation underestimates the fixed-order perturbative uncertainties in the
$0$-jet bin.  In the resummed calculation, fixed-order $\alpha_s$ expansions are
carried out at three distinct scales (hard $\mu_H$, jet/beam $\mu_B$, and soft
$\mu_S$) which appear in the corresponding factorization theorem. The
uncertainties in the resummed cross section are obtained by varying these
scales. Varying $\mu_H$ up and down by a factor of $2$ moves all three scales up
and down, and hence is a scale variation that is correlated with the usual scale
variation for the inclusive cross section. Varying $\mu_B$ or $\mu_S$ while
holding $\mu_H$ fixed explicitly accounts for additional higher order
uncertainties induced by the presence of the large jet-veto logarithms, and hence
allow us to determine $\Delta_\cut$.

In \fig{resum} we compare the remaining perturbative uncertainties after
resummation at NNLL+NNLO, shown by the darker red bands, to the NNLO
uncertainties obtained with the fixed-order method advocated here, which are
shown by the lighter gray bands. The results for the NNLL+NNLO cross section are
obtained from Ref.~\cite{Berger:2010xi}.\footnote{We have made a small
  improvement to Ref.~\cite{Berger:2010xi}. The NNLL+NNLO results of
  Ref.~\cite{Berger:2010xi} fully incorporate the NNLO corrections by adding
  so-called nonsingular fixed-order contributions, which are terms that do not
  appear in an expansion of the strict NNLL result. In Ref.~\cite{Berger:2010xi}
  the nonsingular contributions were obtained for the sum of ${\cal
    O}(\alpha_s)+{\cal O}(\alpha_s^2)$ cross-sections using
  \textsc{FEHiP}~\cite{Anastasiou:2004xq, Anastasiou:2005qj}. Here we use a much higher
  statistics spectrum from \textsc{MCFM}~\cite{Campbell:1999ah}, which allows us to
  separately determine the nonsingular cross sections at ${\cal O}(\alpha_s)$
  and ${\cal O}(\alpha_s^2)$.  The only place this improvement is visible is for
  $\Tcmc\le 3\GeV$, where the resummed cross sections are now consistent with
  zero within the displayed uncertainties.}
The left panel shows the results for the Tevatron and the
right panel the results for the LHC at $7\TeV$. The fact that the
resummation reduces the perturbative uncertainties, as it should, shows that our
method of using independent inclusive scale variations yields more robust
fixed-order uncertainties.

In the resummed calculation, $\sigma_\tot$ is by construction not affected by the
$\mu_S$ and $\mu_B$ variations. We denote the combined $\mu_S$ and $\mu_B$
uncertainty by $\Delta_{SB}$. It provides a direct estimate of the cut-induced
uncertainty, $\Delta_\cut = \Delta_{SB}$, which is anticorrelated between
$\sigma_0(\Tcmc)$ and the corresponding $\sigma_{\geq1}(\Tcmc) = \sigma_\tot -
\sigma_0(\Tcmc)$. On the other hand, the $\mu_H$ variation affects all the cross
sections yielding an uncertainty component that is $100\%$ correlated between them.
In particular, it is responsible for estimating the perturbative
uncertainty of $\sigma_\tot$, for which it is equivalent to the usual
fixed-order scale variation, $\Delta_{H {\rm tot}}=\Delta_\tot$. The full
covariance matrix for $\{\sigma_\tot, \sigma_0,\sigma_{\ge 1}\}$, that is the
analog of \eq{fullmatrix} but for the resummed result, is then
%%%
\begin{align} \label{eq:resummatrix}
C &= C_{SB} + C_H
\,,\nn\\*[1ex]
C_{SB} &=
\begin{pmatrix}
   0 & 0 & 0 \\
   0 & \Delta_{SB}^2 &  - \Delta_{SB}^2 \\
   0 & -\Delta_{SB}^2 & \Delta_{SB}^2
\end{pmatrix},
\\[1ex]\nn
C_H &=
\begin{pmatrix}
   \Delta_{H {\rm tot}}^2 & \Delta_{H {\rm tot}} \Delta_{H0} & \Delta_{H {\rm tot}}\,\Delta_{H\geq 1} \\
   \Delta_{H {\rm tot}}\,\Delta_{H0}  & \Delta_{H0}^2 &  \Delta_{H0}\,\Delta_{H\geq 1}  \\
   \Delta_{H {\rm tot}}\,\Delta_{H\geq 1} & \Delta_{H0}\,\Delta_{H\geq 1} & \Delta_{H\geq 1}^2
\end{pmatrix}
,\end{align}
%%%
where $\Delta_{SB}$ is obtained from the envelope of the $\mu_S$ and $\mu_B$
variations, and $C_{SB}$ is equivalent to $C_\cut$ in \eq{cutmatrix}. The
$\Delta_{Hi}$ are obtained from the $\mu_H$ variation and satisfy $\Delta_{H
  {\rm tot}} = \Delta_{H0} + \Delta_{H\geq 1}$. The full uncertainty in the
$0$-jet bin shown by the darker red bands in \fig{resum} is then given by
$\Delta_{SB}^2+\Delta_{H0}^2$, which is the $0$-bin entry on the diagonal of
$C$.\footnote{In the results of Ref.~\cite{Berger:2010xi}, the envelope of all
  three scale variations was used to obtain the total uncertainty. The slightly
  modified procedure we use here, which adds $\Delta_{SB}$ and $\Delta_{H}$ in
  quadrature, gives very similar results, but has the advantage that it also
  allows for a straightforward treatment of the correlations.}

Compared to \eq{resummatrix}, using a direct exclusive scale variation at fixed order
would correspond to taking $\Delta_{SB}\to 0$ and obtaining the analog of the $\Delta_{Hi}$ by scale
variation without resummation ($\mu_H=\mu_B=\mu_S$). On the other hand, our proposed fixed-order
method would correspond to taking $\Delta_{SB} \to \Delta_{\ge 1}$ and $\Delta_{H\ge
  1} \to 0$, such that $\Delta_{H0} = \Delta_{H{\rm tot}} \to \Delta_\tot$. Hence,
the resummation of the jet-veto logarithms allows one to capture both types of uncertainties appearing in the two different fixed-order methods. Note that the
numerical dominance of $\Delta_{SB}^2$ over $\Delta_{H0}\Delta_{H\ge 1}$ in the
$0$-jet region is another way to justify the
preference for using the combined inclusive scale variation over the direct exclusive scale
variation when given a choice between these two methods.

As an example, consider $\Tcmc = 20\GeV$. At fixed NNLO, the inclusive cross sections are $\sigma_\tot = (8.70 \pm 0.75)\pb$ and $\sigma_{\geq 1} = (2.25 \pm 0.62)\pb$. Using \eq{fullmatrix}, this gives
%%%
\begin{align} \label{eq:Taucorr}
\delta(\sigma_0) &= 15\%
\,,\qquad &
\delta(\sigma_{\geq 1}) &= 28\%
\,,\nn\\
\rho(\sigma_0, \sigma_\tot) &= 0.77
\,,\qquad &
\rho(\sigma_{\geq 1}, \sigma_\tot) &= 0
\,,\nn\\
\rho(\sigma_0, \sigma_{\geq 1}) &= -0.64
\,.\end{align}
%%%
For $\sigma_0$ this corresponds to the lighter gray bands in \fig{resum}, and the structure here is very similar to what we saw in \eq{pTcorr}.

From our resummed result using \eq{resummatrix} we obtain
%%%
\begin{align} \label{eq:Taucorrresum}
\delta(\sigma_0) &= 11.8\%
\,,\qquad &
\delta(\sigma_{\geq 1}) &= 19.7\%
\,,\nn\\
\rho(\sigma_0, \sigma_\tot) &= 0.04
\,,\qquad &
\rho(\sigma_{\geq 1}, \sigma_\tot) &= 0.33
\,,\nn\\
\rho(\sigma_0, \sigma_{\geq 1}) &= -0.82
\,,\end{align}
%%%
which for $\sigma_0$ corresponds to the darker red bands in \fig{resum}.  After
resummation neither of $\sigma_0$ and $\sigma_{\geq 1}$ is strongly correlated
with $\sigma_\tot$ anymore, which at first sight is perhaps a bit surprising.
However, for small $\Tcmc$ this is not unexpected and is simply due to the fact
that the central values and remaining perturbative uncertainties are dominated
by the resummed logarithmic series (i.e. $\Delta_{SB}$ dominates numerically
over $\Delta_{H0}$ and $\Delta_{H\geq 1}$). In fact, this supports our arguments
in \sec{pert}, that the uncertainties from higher-order terms in the logarithmic
series for $\sigma_{\geq 1}$ and the fixed-order series for $\sigma_\tot$ can and
should be considered independent, which lead to \eq{diagmatrix}.

Comparing \eqs{Taucorr}{Taucorrresum}, we see that the uncertainties obtained
from our fixed-order method follow a similar pattern for the relative
uncertainties for $\sigma_0$ and $\sigma_{\geq 1}$ as observed in the resummed
result, with a strong negative correlation between them. Since resummation
provides an improved treatment of the cut-induced effects, we take this as
further evidence that the method of using inclusive fixed-order cross section
uncertainties provides a consistent way to obtain reliable estimates of
perturbative uncertainties in exclusive jet bins. In particular it provides a
suitable starting point for an uncertainty estimate, that can be further refined
when an appropriate resummed result becomes available.

%%%%%%%%%%%%%%%%%%%%%%%%%%%%%%%%%%%%%%%%%%%%%%%%%%%%%%%%%%%%%%%%%%%%%%%%%%%%%%%%
\section{Conclusions}
\label{sec:conclusions}
%%%%%%%%%%%%%%%%%%%%%%%%%%%%%%%%%%%%%%%%%%%%%%%%%%%%%%%%%%%%%%%%%%%%%%%%%%%%%%%%

We have proposed a method to estimate perturbative uncertainties in
fixed-order predictions of exclusive jet cross sections that accounts for the
presence of large logarithms at higher orders caused by the jet binning.
The method uses the
fixed-order calculations of inclusive cross sections, $\sigma_{\geq N}$ and
$\sigma_{\geq N+1}$, for which the standard scale variation provides reasonable
uncertainty estimates, and combines these inclusive uncertainties into an
estimate for the corresponding exclusive $N$-jet cross section $\sigma_N =
\sigma_{\geq N} - \sigma_{\geq N+1}$, treating the inclusive cross sections as
uncorrelated.

We have illustrated this procedure for a variety of processes, including
analysis of $H+0,1$ jets, $WW+0$ jets, and $W+0,1,2$ jets with \textsc{MCFM}, and showed
that it yields more robust estimates of theory uncertainties than direct exclusive scale
variation.  We have also shown for a specific case with $H+0$ jets that it leads
to fixed-order uncertainties that are theoretically consistent with the
corresponding resummed predictions. In jet bins used for new physics searches,
we anticipate that it should yield realistic uncertainty estimates for standard
model backgrounds.  We also expect that it provides a suitable fixed-order
starting point for the central values, uncertainties, and jet bin correlations,
which can be improved by higher-order logarithmic resummation.

Our treatment of the fixed-order exclusive and inclusive cross sections has
followed the standard approach of always using cross section results at the same
order in $\alpha_s$. It would be interesting to study whether this can be
relaxed when using differences of inclusive cross sections to compute the
central values for the jet bins.  For example, for $gg\to H$ one could
independently compute $\sigma_\tot$ at NNLO, and $\sigma_{\geq 1}$ and
$\sigma_{\geq 2}$ each at NLO, and then use these to compute the jet bins as
$\sigma_0 = \sigma_\tot - \sigma_{\geq 1}$ and $\sigma_1 = \sigma_{\geq
  1}-\sigma_{\geq 2}$. Since we argued that the inclusive series can be treated
independently, it may be consistent to include them to different orders to
compute the central value and uncertainties of $\sigma_1$. This would have the
advantage of allowing one to utilize the NLO result for $\sigma_{\geq 2}$
without destroying the consistent perturbative expansion for $\sigma_{\geq 1}$
and $\sigma_\tot$ when the jet bins are added together. Since in this case the
perturbative order of the jet boundary between $\sigma_1$ and $\sigma_{\ge 2}$
does not match up, this deserves a dedicated study before being used in practice.

%%%%%%%%%%%%%%%%%%%%%%%%%%%%%%%%%%%%%%%%%%%%%%%%%%%%%%%%%%%%%%%%%%%%%%%%%%%%%%%%
\begin{acknowledgments}
  We thank the organizers and participants of the workshop ``Higgs at Tevatron
  and LHC'' for stimulating discussion which inspired this work. The workshop
  was sponsored by the University of Washington and supported by the DOE under
  Contract No. DE-FGO2-96-ER40956. We thank Joey Huston, Frank Petriello, and
  Massimiliano Grazzini for stimulating discussions and comments on the manuscript.
  Parts of this work have been carried out within the LHC Higgs Cross
  Section Working Group. This work was supported in part
  by the Office of Nuclear Physics of the U.S.\ Department of Energy under
  Grant No. DE-FG02-94ER40818, and by the Department of Energy under Grant No.
  DE-SC003916.
\end{acknowledgments}
%%%%%%%%%%%%%%%%%%%%%%%%%%%%%%%%%%%%%%%%%%%%%%%%%%%%%%%%%%%%%%%%%%%%%%%%%%%%%%%%

%%%%%%%%%%%%%%%%%%%%%%%%%%%%%%%%%%%%%%%%%%%%%%%%%%%%%%%%%%%%%%%%%%%%%%%%%%%%%%%%
\appendix*
%%%%%%%%%%%%%%%%%%%%%%%%%%%%%%%%%%%%%%%%%%%%%%%%%%%%%%%%%%%%%%%%%%%%%%%%%%%%%%%%

%%%%%%%%%%%%%%%%%%%%%%%%%%%%%%%%%%%%%%%%%%%%%%%%%%%%%%%%%%%%%%%%%%%%%%%%%%%%%%%%
\section{Case of Three Jet Bins}
\label{app:corr}
%%%%%%%%%%%%%%%%%%%%%%%%%%%%%%%%%%%%%%%%%%%%%%%%%%%%%%%%%%%%%%%%%%%%%%%%%%%%%%%%

In this appendix we generalize \eq{fullmatrix} to the case of $0$, $1$, and $(\ge
2)$-jet bins that is actually used in current Higgs searches. Since only neighboring
jet bins will be correlated, the generalization to more than three jet bins is
not any more complicated.

We start from the inclusive cross sections $\sigma_\tot$, $\sigma_{\geq 1}$, $\sigma_{\geq 2}$, and denote their absolute uncertainties by $\Delta_\tot$, $\Delta_{\geq 1}$, $\Delta_{\geq 2}$ and their relative uncertainties by $\delta_i = \Delta_i/\sigma_i$. We define the exclusive cross sections and event fractions
%%%
\begin{align}
\sigma_0 &= \sigma_\tot - \sigma_{\geq 1}
\,,\quad &
f_0 &= \frac{\sigma_0}{\sigma_\tot}
\,,\nn\\
\sigma_1 &= \sigma_\mathrm{\geq 1} - \sigma_{\geq 2}
\,,\quad &
f_1 &= \frac{\sigma_1}{\sigma_\tot}
\,.\end{align}
%%%
The covariance matrix for the four quantities $\{\sigma_\tot, \sigma_0, \sigma_1, \sigma_{\geq 2}\}$ is given by
%%%
\begin{equation} \label{eq:Cov}
C =
\begin{pmatrix}
\Delta_\tot^2 & \Delta_\tot^2 & 0 & 0 \\
\Delta_\tot^2 & \Delta_\tot^2 + \Delta_{\geq 1}^2 & -\Delta_{\geq1}^2 & 0 \\
0 &-\Delta_{\geq1}^2 & \Delta_{\geq 1}^2 + \Delta_{\geq 2}^2 & -\Delta_{\geq2}^2 \\
0 & 0 & -\Delta_{\geq2}^2 & \Delta_{\geq 2}^2
\end{pmatrix}
\,.\end{equation}
%%%
Of course, only three of these four quantities are independent. For example, $\sigma_\tot = \sigma_0 + \sigma_1 + \sigma_{\geq 2}$, and it is easy to check that $\Delta(\sigma_0 + \sigma_1 + \sigma_{\geq 2})^2 = \Delta_\tot^2$, which is given by the sum of all entries in the lower $3\times3$ matrix. The relative uncertainties of $\sigma_{0,1}$ following from \eq{Cov}, written in terms of relative quantities, are
%%%
\begin{align}
\delta(\sigma_0)^2 &= \frac{1}{f_0^2}\, \delta_\tot^2 + \Bigl(\frac{1}{f_0} - 1\Bigr)^2 \delta_{\geq 1}^2
\,,\nn\\
\delta(\sigma_1)^2  &= \Bigl(\frac{1-f_0}{f_1}\Bigr)^2 \delta_{\geq 1}^2 +
\Bigl(\frac{1-f_0}{f_1} - 1\Bigr)^2 \delta_{\geq 2}^2
\,.\end{align}
%%%
Similarly, the correlation coefficients for $\sigma_{0}$ and $\sigma_1$ following from \eq{Cov} are
%%%
\begin{align}
\rho(\sigma_0, \sigma_\tot)
&= \biggl[1 + \frac{\delta_{\geq 1}^2}{\delta_\tot^2}(1-f_0)^2 \biggr]^{-1/2}
\,,\nn\\
\rho(\sigma_0, \sigma_1)
&= - \biggl[1 + \frac{\delta_\tot^2}{\delta_{\geq 1}^2}\, \frac{1}{(1-f_0)^2}\biggr]^{-1/2}
\nn\\ & \quad\times
\biggl[1 + \frac{\delta_{\geq2}^2}{\delta_{\geq 1}^2} \Bigl(1 - \frac{f_1}{1-f_0}\Bigr)^2\biggr]^{-1/2}
\,,\nn\\
\rho(\sigma_0, \sigma_{\geq 2}) &= 0
\,,\nn\\
\rho(\sigma_1, \sigma_\tot) &= 0
\,,\nn\\
\rho(\sigma_1, \sigma_{\geq 2})
&= -\biggl[1 + \frac{\delta_{\geq 1}^2}{\delta_{\geq2}^2} \Bigl(1 - \frac{f_1}{1-f_0}\Bigr)^{-2} \biggr]^{-1/2}
\,.\end{align}
%%%
The relative uncertainties for $f_0$ and $f_1$ are
%%%
\begin{align}
\delta(f_0)^2
&= \Bigl(\frac{1}{f_0} - 1 \Bigr)^2\bigl(\delta_\tot^2 + \delta_{\geq 1}^2\bigr)
\,,\\\nn
\delta(f_1)^2
&= \delta_\tot^2 + \Bigl(\frac{1-f_0}{f_1}\Bigr)^2 \delta_{\geq 1}^2 +
\Bigl(\frac{1-f_0}{f_1} - 1\Bigr)^2 \delta_{\geq 2}^2
\,,\end{align}
%%%
and their correlations are
%%%
\begin{align}
\rho(f_0, \sigma_\tot)
&= \biggl[1 + \frac{\delta_{\geq 1}^2}{\delta_\tot^2} \biggr]^{-1/2}
\,,\nn\\
\rho(f_0, f_1)
&= -\biggl(1 + \frac{1-f_0}{f_1}\frac{\delta_{\geq 1}^2}{\delta_\tot^2} \biggr)
\Bigl(\frac{1}{f_0} - 1\Bigr) \frac{\delta_\tot^2}{\delta(f_0)\delta(f_1)}
\,,\nn\\
\rho(f_1, \sigma_\tot)
&= -\frac{\delta_\tot}{\delta(f_1)}
\,.\end{align}
%%%

\bibliographystyle{../../physrev4}
\bibliography{../../pp}

\end{document}